\documentclass[journal]{IEEEtran}
\ifCLASSINFOpdf
\else
\fi
\usepackage{float}
\usepackage{amssymb}
\usepackage{cite}
\usepackage{amsmath,amssymb,amsfonts}
\usepackage{algorithmic}
\usepackage{graphicx}
\usepackage{textcomp}
\usepackage{xcolor,soul}
\usepackage[english]{babel}
\usepackage{amsthm}
\usepackage[english]{babel}
\usepackage{verbatim}
\usepackage{comment}
\newtheorem{theorem}{Theorem}

\newtheorem{assumption}{Assumption}
\newtheorem{proposition}{Proposition}
\newtheorem{definition}{Definition}

\usepackage[ruled]{algorithm2e}
\usepackage{ulem}
\usepackage{pstricks}
\usepackage{xurl}
\usepackage{bm}
\usepackage{bbm}
\usepackage{mathtools}

\DeclareFontFamily{OT1}{pzc}{}
\DeclareFontShape{OT1}{pzc}{m}{it}{ <-> s*[1.15] pzcmi7t }{}
\DeclareMathAlphabet{\mathpzc}{OT1}{pzc}{m}{it}

\allowdisplaybreaks

\newcommand{\eps}{\varepsilon}

\DeclareMathOperator*{\minimize}{minimize}

\newcommand\bbE{\ensuremath{\mathbb{E}}} 

\newcommand{\mat}[1]{\boldsymbol{#1}}
\renewcommand{\vec}[1]{\boldsymbol{\mathrm{#1}}}
\newcommand{\vecalt}[1]{\boldsymbol{#1}}

\newcommand{\normof}[1]{\|#1\|}

\newcommand{\abs}[1]{\vert #1 \vert}

\newcommand{\RR}{\mathbb{R}}

\newcommand{\LL}{\mathcal{L}}

\newcommand{\inner}[2]{\langle #1, #2 \rangle}

\providecommand{\mD}{\ensuremath{\mat{D}}}

\providecommand{\vx}{\ensuremath{\vec{x}}}

\providecommand{\vpi}{\ensuremath{\vecalt{\pi}}}

\providecommand{\vlambda}{\ensuremath{\vecalt{\lambda}}}
\providecommand{\Ttrain}{\ensuremath{T_\text{train}}}
\providecommand{\Thour}{\ensuremath{T_\text{hour}}}
\providecommand{\Tday}{\ensuremath{T_{\text{day}}}}
\providecommand{\vbelief}{\ensuremath{\hat{\vlambda}}}

\providecommand{\nd}{\ensuremath{q}}
\providecommand{\reg}{\text{REG}}

\providecommand{\vLL}{\ensuremath{\vec{\LL}}}

\begin{document}
\title{DER Aggregators in Wholesale Electricity Markets: A Hybrid Mean-Field Framework with Endogenous Price Feedback}

\author{Jun~He 
        and~Andrew~L.~Liu
\thanks{Jun He is with Edwardson School of Industrial Engineering, Purdue University, West Lafayette, IN, USA, email: he184@purdue.edu.}
\thanks{Andrew L. Liu is with Edwardson School of Industrial Engineering, Purdue University, West Lafayette, IN, USA, email: andrewliu@purdue.edu.}}

%
%

\markboth{Journal of \LaTeX\ Class Files,~Vol.~14, No.~8, August~2015}%
{Shell \MakeLowercase{\textit{et al.}}: Bare Demo of IEEEtran.cls for IEEE Journals}
%



\maketitle

\begin{abstract}
The rapid growth of distributed energy resources (DERs), including rooftop solar and energy storage, is transforming the grid edge, where customer-side assets increasingly interact with the broader power grid. DER aggregators, entities that coordinate the actions of many small-scale DERs, play a key role in this transformation. This paper presents a hybrid Mean-Field Control (MFC) and Mean-Field Game (MFG) framework for integrating DER aggregators into wholesale electricity markets. Unlike traditional approaches that treat market prices as exogenous, our model captures the feedback between aggregators’ collective actions and electricity prices. The MFC component optimizes DER operations within each aggregator, while the MFG models strategic interactions among multiple aggregators. To account for various uncertainties, we incorporate reinforcement learning, which allows aggregators to learn optimal bidding strategies in dynamic market conditions. We prove the existence and uniqueness of a mean-field equilibrium and validate the framework through a case study of the Oahu Island power system. Results show that our approach reduces price volatility and improves market efficiency, offering a scalable and decentralized solution for DER integration in wholesale markets.

\end{abstract}


%
%
%
%

\section{Introduction}
\label{sec:intro}
DERs, including rooftop solar and energy storage, are typically located behind the meter; that is, on the customer’s side of the electricity meter, where energy is generated, stored, or managed locally rather than delivered by a utility company. As a result, end-users are evolving from passive consumers into prosumers who can both produce and consume electricity. Prosumers are becoming increasingly active in grid operations through bi-directional power flows, demand flexibility, and localized generation, collectively transforming the traditionally centralized grid into a more distributed and participatory system. 

Recognizing the significance of this shift, the Federal Energy Regulatory Commission (FERC) issued Order 2222 in 2020 to enable distributed energy resources to participate in wholesale electricity markets. However, individual DERs are typically too small to meet minimum size or performance thresholds. In this context, aggregators, also known as virtual power plants (VPPs), are increasingly serving as intermediaries. In practice, companies like OhmConnect aggregate over 200,000 households to bid into CAISO as virtual resources \cite{OhmConnect}, while Tesla has deployed aggregated Powerwall battery systems in ERCOT \cite{Tesla}.  
Yet, integrating large numbers of heterogeneous, small-scale resources into a wholesale market introduces new challenges in understanding the market outcomes of their decentralized participation, including impacts on price volatility and system reliability.

Current research has extensively examined the strategies of how a single aggregator manages DER portfolios, often under the simplifying assumption that wholesale market prices, such as locational marginal prices (LMPs), are exogenous (for example, \cite{contreras2017participation,Birge,RobustBidding}).  While this perspective offers valuable insights into aggregator operations, it neglects an important aspect of market integration; that is, aggregators’ collective actions can influence LMPs. In practice, as aggregators participate in wholesale markets at scale, their decisions contribute to price formation, making the system inherently more coupled and dynamic. Modeling this price feedback is essential for understanding the full impact of DER integration. 

To address this need, we propose a prescriptive, learning-based modeling framework grounded in MFG theory. Rather than describing current market behavior, our approach aims to engineer a scalable and decentralized framework in which aggregators learn to optimize their DER portfolios in response to market prices shaped by the collective behavior of others. Using reinforcement learning (RL), each aggregator adapts its strategy over time based on observed market feedback, enabling dynamic decision-making in uncertain and evolving conditions. We establish theoretical guarantees by proving the existence and uniqueness of a mean-field equilibrium (MFE), which characterizes the steady-state outcome when aggregators adopt policies learned through this process. By explicitly modeling the feedback between aggregator decisions and market prices, and enabling each aggregator to learn optimal bidding strategies in a decentralized and scalable way, our framework lays the groundwork for AI-enabled, automated control architectures that support real-time, market-integrated coordination of DERs.

While the MFG framework captures interactions among aggregators at the market level, each aggregator still faces the challenge of managing a large number of DERs internally. To ensure scalability and tractability, we employ a MFC approach to optimize the operation of each aggregator’s DER portfolio. A key difficulty lies in the presence of energy storage, which links decisions across time, and various uncertainties, including renewable intermittency, load fluctuations, and market price variability, whose distributions are often unknown or difficult to model accurately. Traditional optimization-based methods struggle in this setting due to their reliance on full system observability, centralized coordination, or precise probabilistic modeling. In contrast, our approach combines MFC with RL to learn optimal control policies from interaction with the environment, without requiring prior knowledge of system dynamics or uncertainty distributions. By solving for an optimal policy for a representative DER that reflects the population’s aggregate behavior, this framework significantly reduces computational complexity and enables each aggregator to learn adaptive, forward-looking strategies.


The contributions of this paper are threefold:
(1) We develop a hybrid MFC and MFG framework, integrated with RL, that enables decentralized DER aggregators to participate in wholesale electricity markets under various uncertainties. 
(2) We establish theoretical guarantees for the system by proving the existence and uniqueness of an MFE under entropy-regularized RL, providing theoretical guarantees for well-posedness and consistency of the system behavior in the infinite-agent limit.
(3) We implement a two-phase RL algorithm for the hybrid MFC-MFE model and demonstrate its effectiveness in a case study based on the Oahu Island power system. The results show that it stabilizes aggregator behavior and market prices, reduces price volatility, and lowers system-wide costs, demonstrating the potential of our approach for AI-enabled coordination at the grid edge.

A central contribution of this work is to provide a scalable, stable, and fully decentralized market implementation whose convergence is theoretically guaranteed. This matters directly for market design: ISO/RTOs cannot issue dispatch instructions to behind-the-meter DERs, and the full centralized stochastic dispatch problem, with energy storage, it is bound to be a mutli-period problem, to accounting for millions of DERs and multi-period uncertainties, cannot be solved in practice by any existing optimization algorithm or hardware. Even for research-scale systems, only a severely simplified and approximate version of the centralized problem can be solved numerically; this approximation is used in our paper solely to provide a performance reference, not as a feasible or implementable market solution.
Within this context, our contribution is to show that the proposed decentralized mean-field (MF) framework converges to a stable equilibrium with no scaling bottleneck: the ISO/RTO continues to run its standard unit commitment (UC) or economic dispatch (ED), while DER coordination occurs locally through aggregators with guaranteed convergence and no central-state explosion.
Although a formal price-of-anarchy bound remains an open direction, our results move toward that goal: the decentralized equilibrium exhibits performance nearly indistinguishable from the (approximated) centralized reference, and significantly outperforms other decentralized or MARL-based approaches that lack equilibrium guarantees or often fail to converge. These findings provide both a practically implementable coordination architecture for large-scale DER integration and concrete insights for future market design.

The rest of the paper is organized as follows. Section \ref{sec:lit} reviews the related literature, while Section \ref{sec:Wholesale} presents a baseline electricity market model, including the mathematical formulation and a key property of LMPs. Section \ref{sec:Agg} formulates the aggregators’ decision-making problem using the hybrid MFC–MFG framework. Section \ref{sec:mfe} proves the existence and uniqueness of a MFE and proposes a two-phase distributed learning algorithm. Section \ref{sec:numerical} presents numerical results based on the Oahu Island power system to demonstrate system-level benefits. Finally, Section \ref{sec:conclusion} concludes the paper and outlines directions for future research.
\section{Literature Review} \label{sec:lit}
The literature on bidding strategies in wholesale electricity markets spans extensive work on both single-agent and multi-agent scenarios, using either optimization-based or learning-based methodologies. For single-agent bidding, notable contributions include \cite{IRIA20191361}, which formulates a two-level optimization model, and \cite{ye-ddpg-single-bidding}, which applies deep RL techniques. Similar methodological directions arise in the multi-agent context. For instance, \cite{fallahi2019linear} leverages linear programming to address multi-agent demand response bidding, while \cite{shafie2014stochastic} introduces a multilayer stochastic agent-based model to examine wholesale market dynamics with renewable energy resources. Additionally, \cite{chen2024wholesale} proposes an auction framework that facilitates aggregator participation in wholesale markets without necessitating real-time intervention from a distribution system operator. Learning-based approaches, in contrast, are predominantly applied to peer-to-peer energy trading markets. For example, \cite{liu2021multi} employs multi-agent RL (MARL) with multi-armed bandits in repeated auction scenarios for peer-to-peer energy transactions.

While MARL is conceptually connected to game-theoretic equilibria such as Nash equilibria, exact convergence is rarely guaranteed, particularly in stochastic games. Moreover, as the number of agents increases, equilibrium computation becomes intractable \cite{zhang2021multi}. To address this limitation, recent research has turned to the concept of MFE, which is defined for games with an infinite population. In this framework, the multi-agent problem reduces to the interaction between a representative agent and the distribution of the population’s average state -- the mean field. This formulation greatly enhances computational tractability and serves as an approximation of agent behavior in large but finite populations.

Various studies have explored mean-field frameworks in different settings. For scenarios where system dynamics are known, \cite{feng2025decentralized} employs a dynamic programming approach in response to the mean field. Conversely, in cases where system dynamics are unknown, RL has been applied. For example, \cite{guo2019learning} develops a general MFG framework using RL under competitive conditions, whereas \cite{xie21g-lwp} incorporates fictitious play within MFG to enable agents to learn during gameplay. Whereas most literature assumes the availability of a mean-field oracle, \cite{sandbox}  outlines a Sandbox learning algorithm for situations where mean-field information is not directly accessible to agents. Finally, \cite{mondal2023mean} investigates the use of MFC for approximating MARL problems involving cooperative, heterogeneous agents.
\section{Wholesale Market Model and Locational Marginal Prices}\label{sec:Wholesale}
To ensure the paper is self-contained and accessible to a broad audience, this section presents a standard wholesale electricity market model along with the formulation of LMPs. 

Consider a wholesale electricity market operating over a transmission network with $N$ buses, $L$ transmission lines, and $G$ bulk generators. Each bus $n \in \mathcal{N}:=\{1, \ldots, N\}$ serves a group of $M_n$ agents, comprising $M_n^p$ prosumers and $M_n^c$ pure consumers without any DERs. Each bulk generator $g \in \mathcal{G}:=\{1, \ldots, G\}$ is characterized by a cost function $C_g(\cdot)$ that captures the cost of electricity production. Let $G_n \subseteq \mathcal{G}$ denote the set of generators connected to bus $n$. We require the sets to be disjoint and collectively exhaustive; that is,  $\cup_{n=1}^N G_n = \mathcal{G}$ and $G_n \cap G_{n'} = \emptyset$ for any $n \ne n'$. The transmission lines are indexed by $l \in \mathcal{L} := \{1, \dots, L\}$.

On the demand side, let \( d^n_{it} \) denote the net demand of prosumer \( i \) at bus \( n \), defined as total energy consumption minus local solar generation. If \( d^n_{it} > 0 \), the prosumer is a net consumer; if \( d^n_{it} < 0 \), the prosumer exports energy to the grid. This bidirectional interaction allows prosumers to participate flexibly in the energy market. Similarly, we use \( d^n_{jt} \geq 0 \) to denote the demand of consumer \( j \) at bus \( n \). The total demand at bus \( n \) and time \( t \) is:
\begin{equation}
    D^n_t = \sum_{i=1}^{M^p_n} d^n_{it} + \sum_{j=1}^{M^c_n} d^n_{jt}. \label{eq:Demand_n}
\end{equation}
To avoid technical complications from infeasible supply-demand imbalances, we assume that total net demand is non-negative at each timestep; that is, $
    \sum_{n=1}^{N} D^n_t \geq 0,\ t = 1, 2, \ldots.
$
This is reasonable under current DER penetration levels, where distributed generation does not yet exceed total system demand.

The wholesale market operates over discrete time intervals, indexed by \( t \in \{1, 2, \ldots\} \), where each timestep corresponds to a fixed duration, such as an hour. The duration of each time interval is denoted by $\Delta t$.
During each timestep, the ISO of a wholesale electricity market collects aggregate supply and demand bids and solves an optimization problem to determine the least-cost dispatch that satisfies system constraints. The marginal cost of supplying one additional unit of electricity at each node, exactly the LMPs, is derived from the dual variables associated with the power balance and transmission constraints.

To provide the specific formula of LMPs, we present a simplified version of the ED problem solved by an ISO. Although actual dispatch problems are considerably more complex, this  formulation retains the essential constraints: power balance, generation capacity, and transmission limits, and is sufficient to demonstrate how prices are formed and how they relate to system demand. Importantly, our framework is not restricted to this simplification. A key strength of the framework is that it requires no modification to ISO operations: system operators can continue using their full-scale market-clearing processes without adjustment. All learning and coordination occur at the aggregator level, ensuring full compatibility with existing market structures.

The simplified ED problem is given as follows:
\begin{align}
    & \min_{p_{1t}, \cdots, p_{Gt}} \; \sum_{g=1}^G C_g(p_{gt}) \label{eqn:ed-obj} \\
    & \text{s.t.} \; \sum_{g=1}^G p_{gt} = \sum_{n=1}^{N} \frac{D^n_t}{\Delta t}, \quad (\lambda^{\text{HUB}}_t) \label{eqn:power-balance} \\
    & \quad -\overline{F}_l \leq \sum_{n=1}^{N} \text{PTDF}_{ln} \left( \sum_{g \in G_n} p_{gt} - \frac{D^n_t}{\Delta t} \right) \leq \overline{F}_l,\ (\underline{\mu}_{lt}, \overline{\mu}_{lt}) \nonumber \\
    & \hspace{180pt}\quad \forall \; l \in \mathcal{L}, \label{eqn:flow-limit} \\
    & \quad 0 \leq p_{gt} \leq \overline{p}_g, \quad (\underline{\nu}_{gt}, \overline{\nu}_{gt}) \quad \forall \; g \in \mathcal{G}. \label{eqn:gen-limit}
\end{align}
Here, the variables $p_{gt}$ represent the average power output of generator $g$ in time interval $t$. The parameter $\text{PTDF}_{ln}$ denotes the power transfer distribution factor for line $l$ and bus $n$, which reflects how power flows across the network; $\overline{F}_l$ represents the maximum allowable power flow on line $l$, and $\overline{p}_g$ represents the generation capacity of generator $g$. In addition, the dual variables are listed on the right-hand-side of each constraint. The dual of the power balance constraint \eqref{eqn:power-balance}, $\lambda^{\text{HUB}}_t$, represents the so-called hub price, and $\underline{\mu}_{lt}, \overline{\mu}_{lt}, \underline{\nu}_{gt}, \overline{\nu}_{gt}$ are the dual variables corresponding to line flow limits of each line $l$, and generator output limits of each generator $g$, respectively. The LMP at bus $n \in \mathcal{N}$ and time $t$, denoted $\lambda^n_t$, is derived as follows:
\begin{align}
    \lambda^n_t & \coloneq \frac{\partial \mathfrak{L}_t}{\partial D^n_t} = \lambda^{\text{HUB}}_t - \sum_{l=1}^L \text{PTDF}_{ln} (\overline{\mu}_{lt} - \underline{\mu}_{lt}), 
    \label{eqn:lmp}
\end{align}
where $\mathfrak{L}_t$ is the Lagrangian function of the ED problem.  For notation simplicity, let $\mD_t := (D^1_t, \cdots, D^N_t)$ denote the vector of all demand bids; then the LMP $\lambda^n_t$ is a function of $\mD_t$, which we write as $\lambda^n_t(\mD_t)$.

In establishing the existence and uniqueness of an MFE in later sections, a key technical condition needed is the Lipschitz continuity of the mapping \( \lambda^n_t(\mD_t) \).  The following result, established in \cite{feng2025decentralized}, provides sufficient conditions under which this property holds. For completeness, we restate it here, along with the required constraint qualification assumption.

\begin{assumption}[LICQ] \label{assump:LICQ}
    Let \( X(\mD_t) \) denote the feasible region of the economic dispatch problem defined in \eqref{eqn:ed-obj}–\eqref{eqn:gen-limit}. We assume that, for all \( t \) and all \( \mD_t \) such that \( X(\mD_t) \) is non-empty, the linear independence constraint qualification (LICQ) holds at every feasible point in \( X(\mD_t) \).
\end{assumption}

\begin{proposition}[Lipschitz Continuity of LMPs\cite{feng2025decentralized}] 
    Given Assumption \ref{assump:LICQ} and that each generator’s cost function \( C_g(\cdot) \) is convex and quadratic. Then, for all buses \( n = 1, \ldots, N \), the LMP \( \lambda^n_t(\mD_t) \) is single-valued and Lipschitz continuous with respect to \( \mD_t \); that is, there exists a constant \( L_{\lambda} > 0 \) such that for any feasible $\mD_t$ and $\widetilde{\mD}_t \geq 0$, $
        \abs{\lambda^n_t(\mD_t) - \lambda^n_{t}(\widetilde{\mD}_t)} \leq L_{\lambda} \normof{\mD_t - \widetilde{\mD}_t}_1.
$
    \label{prop:feng2025}
\end{proposition}
\section{The MF Framework of a Single Aggregator}
\label{sec:Agg}
In this section, we formulate the individual decision-making problem faced by a single aggregator. Acting on behalf of a large population of prosumers equipped with solar PVs and energy storage, the aggregator must determine control strategies for the prosumers under multiple sources of uncertainty, including solar energy output, real-time energy demand, and market prices. A key challenge lies in the presence of energy storage, which links decisions across time and leads to a high-dimensional stochastic dynamic optimization problem. To address this complexity in a scalable manner, especially since the aggregator typically manages a large number of prosumers, we adopt an MFC approach, which approximates the collective behavior of prosumers using a representative agent and enables decentralized learning of control policies.

In power systems, many processes exhibit strong intra-day cycles that also repeat with similar patterns across days. For example, electricity demand and solar generation vary significantly across hours of the day, yet their daily profiles tend to follow recurring diurnal patterns. To capture both intra-day variation and inter-day similarity, we define two mappings over the global timestep $t = 0, 1, 2, \ldots$. Let $H$ denote the number of timesteps in a day, then
\begin{equation}\label{eq:time_map}
\Thour(t) = (t \bmod H) + 1 \quad \text{and} \quad \Tday(t) = \left\lfloor \frac{t}{H} \right\rfloor,
\end{equation}
which return the hour within the day and the day index, respectively. These mappings enable the model to encode hourly dynamics while treating each day as structurally similar.

Before introducing the MF structure, we first present the aggregator's problem in a standard reinforcement learning (RL) form, i.e., as an infinite-horizon discounted Markov decision process with state, action, transition, and reward. We then connect this RL formulation to MFC by noting that the LMP can be modeled as a function of the MF distribution over state-action pairs. Notation-wise, we use $x \sim \mathcal{Q}$ to indicate that $x$ follows the distribution $\mathcal{Q}$; we use $\mathcal{P}(\mathcal{X})$ to be the set of all Borel probability measures on any space $\mathcal{X}$; we use $\abs{\mathcal{X}}$ to denote the cardinality of $\mathcal{X}$ if it is finite and discrete; and we use $[ \vx ]_k$ to denote the $k$-th entry of the vector $\vx$.

\subsection{Reinforcement Learning}
We first describe the aggregator's decision problem using the standard RL framework. In this formulation, the aggregator is viewed as an agent that interacts repeatedly with a stochastic environment and learns a policy that maximizes a long-term performance criterion. At each decision epoch $t$, the agent observes the current system state $s_t$, selects an action $a_t$ according to a policy, and receives a scalar reward that reflects the immediate economic outcome of this decision. The system then transitions to a new state $s_{t+1}$ according to the underlying system dynamics and exogenous uncertainties. By iteratively interacting with the environment over time, the aggregator aims to learn a policy that maximizes the expected discounted sum of rewards over an infinite horizon.

The only decisions in our setup involve charging or discharging aggregator's energy storage. Let $\mathcal{A} \subseteq [-1, 1]$ denote the action space, where each action $a_t \in \mathcal{A}$ represents the proportion of storage capacity to charge (if $a_t > 0$) or discharge (if $a_t < 0$) at time $t$, drawn according to the policy learned by the aggregator and executed by individual prosumers. To ensure feasibility, we implement \textit{action masking}, a common technique in RL that eliminates invalid actions based on the current state. Specifically, the algorithm assigns a value of $-\infty$ to infeasible actions, resulting in zero probability of selection. 

The aggregator's state is characterized by three state variables at time $t$: (i) storage level $e_t \in [0,1]$ (as a percentage of capacity), (ii) net load $\nd_{t}$ (unrelated to storage charging/discharging), and (iii) the current hour of the day, $\Thour(t)$. Let the state space be $\mathcal{S} \subseteq \RR^3$, where each element is a tuple of the form $s_{t} := (e_{t}, \nd_{t}, \Thour(t)) \in \mathcal{S}$. Among the state variables, the net load $\nd_{t}$ is an exogenous random variable, mainly driven by solar irradiance and consumption behavior, and is unaffected by aggregators' actions. Time of day is deterministic. The only state variable affected by decisions is the storage level, which evolves according to the following rule after action $a_{t}$:
\begin{equation}
    e_{t+1} := \max \{\min \{e_{t} + a_{t}, 1\}, 0 \}.
    \label{eqn:storage-transition}
\end{equation}

We now define the single-period reward function $r: \mathcal{S} \times \mathcal{A} \times \RR^H \to \RR$ for the aggregator. Let $s_t$ and $a_t$ be the state and action at time $t$, and let $\vlambda := (\lambda_1, \dots, \lambda_H)$ denote the LMP profile. The reward is given by
\begin{equation}
    r(s_t, a_t, \vlambda) = - \lambda_{\Thour(t)} \cdot \overline{E} \cdot \big( \Phi(e_t, a_t, \eta) + \nd_t \big),
    \label{eqn:rew}
\end{equation}
where $\overline{E}$ is the aggregator's storage capacity, $\eta \in (0, 1]$ is the storage efficiency (assumed uniform across prosumers at bus $n$), and $\Phi(\cdot)$ adjusts the action for efficiency losses:
\begin{equation}\label{eqn:adjusted-action}
   \Phi(e, a, \eta) = 
   \begin{cases}
   \max \{-e, a \} \cdot \eta, & \text{if } a < 0\ \text{(discharging)}, \\
   \min \{1 - e, a \} / \eta, & \text{if } a \geq 0\ \text{(charging)}. 
   \end{cases}
\end{equation}
This adjustment accounts for efficiency losses during storage operations. For instance, if $a > 0$ and $\min\{1 - e, a\} = a$, then to increase the storage level by $a$ (as a percentage of total capacity), the aggregator must purchase $a / \eta$ worth of energy from the grid to achieve the desired charge level, indicating that storage inefficiency requires drawing more energy than is ultimately stored. The reward represents net profit, computed as negative price multiplied by total energy demand. A positive total quantity, comprising net load $\nd_t$ and charging/discharging $\Phi$, indicates a net energy purchase (resulting in a negative reward), while a negative total quantity reflects energy sold back to the grid (resulting in a positive reward). Since $r$ is linear in the LMP profile $\vlambda$, its Lipschitz constant with respect to $\vlambda$ can be read directly from the reward definition. In particular, the net quantity $\Phi(e,a,\eta)+q$ is uniformly bounded over $\mathcal{S}\times\mathcal{A}$. We define
\begin{equation}
    L_r \coloneq \overline{E} \max_{(s,a)\in\mathcal{S}\times\mathcal{A}} \abs{\Phi(e,a,\eta)+q}. 
    \label{eqn:reward-lipschitz-constant}
\end{equation}
Then, for any $(s,a)\in\mathcal{S}\times\mathcal{A}$ and any two LMP profiles $\vlambda_1,\vlambda_2\in\RR^H$,
\begin{equation}
    \abs{r(s,a,\vlambda_1)-r(s,a,\vlambda_2)} \leq L_r\normof{\vlambda_1-\vlambda_2}_1.
    \label{eqn:reward-lipschitz}
\end{equation}

At each decision time $t$, the aggregator learns a policy  $\pi_t: \mathcal{S} \times \RR^H \to \mathcal{P}(\mathcal{A})$, which maps each state and LMP profile to a distribution over actions. Given the LMP profile $\vlambda \in \RR^H$ and initial state $s_0$, the value of policy $\pi_t$ is defined as the expected sum of discounted rewards:
\begin{equation*}
V(s_0, \pi_t, \vlambda) = \bbE \left[ \sum_{\tau=0}^{\infty} \gamma^\tau r(s_{\tau}, a_{\tau}, \vlambda) \right],
\end{equation*}
where $\gamma \in [0, 1)$ is the discount factor, and the expectation is over the trajectory induced by the policy and external uncertainties. To promote exploration and improve learning stability, we incorporate \textit{entropy regularization} into the reward function. Specifically, the single-period reward is modified as:
\begin{equation}\label{eq:r_Reg}
r^\reg(s, a, \vlambda^n) = r(s, a, \vlambda) - \Omega(\pi(\cdot \mid s, \vlambda)),
\end{equation}
where $\Omega(\cdot)$ is a $\rho$-strongly convex regularization function. A common choice is the negative entropy:
\begin{equation*}
\Omega(\pi(\cdot \mid s, \vlambda^n)) = \alpha \sum_{a \in \mathcal{A}} \pi(a \mid s, \vlambda) \log \pi(a \mid s, \vlambda),
\end{equation*}
with $\alpha > 0$ controlling the strength of regularization. The corresponding \textit{regularized value function} is:
\begin{equation}\label{eq:V_Reg}
V^\reg(s_0, \pi_t, \vlambda) = \bbE \left[ \sum_{\tau=0}^{\infty} \gamma^\tau r^\reg(s_{\tau}, a_{\tau}, \vlambda) \right].
\end{equation}
The aggregator seeks to learn an optimal policy $\pi^*_t$ that maximizes $V^\reg$ at each time $t$.

\subsection{An MFC Interpretation of the Learning Problem}
The aggregator represents a large population of $M^p$ prosumers equipped with solar PV and energy storage, together with $M^c$ pure consumers. The total aggregated storage capacity managed by the aggregator is capped at $\overline{E}$. To capture heterogeneity while admitting a MF limit, prosumers are partitioned into $K$ types indexed by $k \in \{1,2,\ldots,K\}$, with proportions $b_k \in [0,1]$ satisfying $\sum_{k=1}^K b_k = 1$. Each type-$k$ prosumer is assigned a relative storage capacity parameter $\theta_k > 0$, so that the individual storage capacity of a type-$k$ prosumer $i$ is given by
\begin{equation*}
    \overline{E}_{ik} := \frac{\theta_k \overline{E}}{M^p \sum_{\kappa=1}^K \theta_\kappa b_\kappa}.
\end{equation*}
This scaling ensures that, as the number of prosumers grows large, the aggregate storage capacity remains finite. Although real systems consist of finitely many agents, the MF approximation becomes accurate when the population is sufficiently large, since the impact of any single agent on aggregate outcomes becomes negligible. In practice, additional heterogeneity (\textit{e.g.}, differing risk preferences, battery degradation costs, or stochastic usage patterns) can be incorporated by increasing the number of agent classes or by parameterizing the policy network with user-specific features. This affects computational scale but does not alter the equilibrium structure (see later sections).

Given these individual capacities, each prosumer's net demand is expressed in normalized form. At time $t$, let $\nd_{it}$ denote the ratio of prosumer $i$'s net demand to its storage capacity $\overline{E}_i$, so that $\nd_{it} \in [-1,1]$. Net demand is inherently stochastic, and we denote its (unknown) distribution by $\mathcal{Q}^p$. Neither prosumers nor aggregators are assumed to know $\mathcal{Q}^p$ a priori; instead, optimal behavior is learned from data through repeated interaction with the environment. For pure consumers, we introduce an analogous normalization using a reference capacity $\overline{E}_j$, yielding normalized demand $\nd_{jt}$ with distribution $\mathcal{Q}^c$.

Rather than having each prosumer learn an individual policy, which is impractical due to market participation constraints such as the 100~kW minimum threshold in FERC Order No.~2222, the aggregator acts as the decision-making agent on behalf of the entire population. Since prosumers at the same location face identical external conditions (e.g., prices and weather), a common control policy is justified. In our setup, behind-the-meter solar generation is first used to meet local demand which will be counted towards net load; storage changes through the selected charging or discharging action. The aggregator's only control decision concerns additional charging from the grid or discharging storage to sell energy back to the market. At each time $t$, the aggregator learns and updates a policy mapping observed system states to charging or discharging actions, which is then executed uniformly across the population.

Up to this point, the aggregator's problem can be viewed as a standard RL problem with a given price signal. The key distinction in our setting is that market prices are not exogenous. Instead, the LMP profile is determined endogenously by system-level market clearing, which depends on the aggregate behavior of all agents in the system. As a result, the reward received by the aggregator depends on the population only through the collective distribution of states and actions. This observation leads naturally to a MFC formulation, in which individual optimization is coupled through an aggregate MF. 

Formally, when the number of prosumers becomes large, aggregate behavior is characterized by the distribution of state-action pairs. For simplicity, we assume that the state space $\mathcal{S}$ and action space $\mathcal{A}$ are discrete and finite, and let $\mathcal{P}(\mathcal{S} \times \mathcal{A})$ denote the set of probability measures over $\mathcal{S} \times \mathcal{A}$. The MF at time $t$ is defined as $\mathcal{L}_t \in \mathcal{P}(\mathcal{S} \times \mathcal{A})$, where for any $s \in \mathcal{S}$ and $a \in \mathcal{A}$:
\begin{equation}\label{eq:MF}
\LL_t(s,a):=\lim_{M^p\to\infty}\frac{1}{\sum_{i=1}^{M^p}\overline E_i} \sum_{i=1}^{M^p}\overline E_i \mathbbm{1}_{(s_{it},a_{it})=(s,a)}.
\end{equation}
Intuitively, the MF captures the limiting empirical distribution of states and actions across the population. Through market clearing, this distribution determines the LMP profile faced by the aggregator, thereby closing the feedback loop between individual reinforcement learning decisions and aggregate system outcomes.
\section{Multiple Aggregators' Problem and the MFE}\label{sec:mfe}
\begin{figure}[ht]
\centering
\includegraphics[width=0.8\columnwidth]{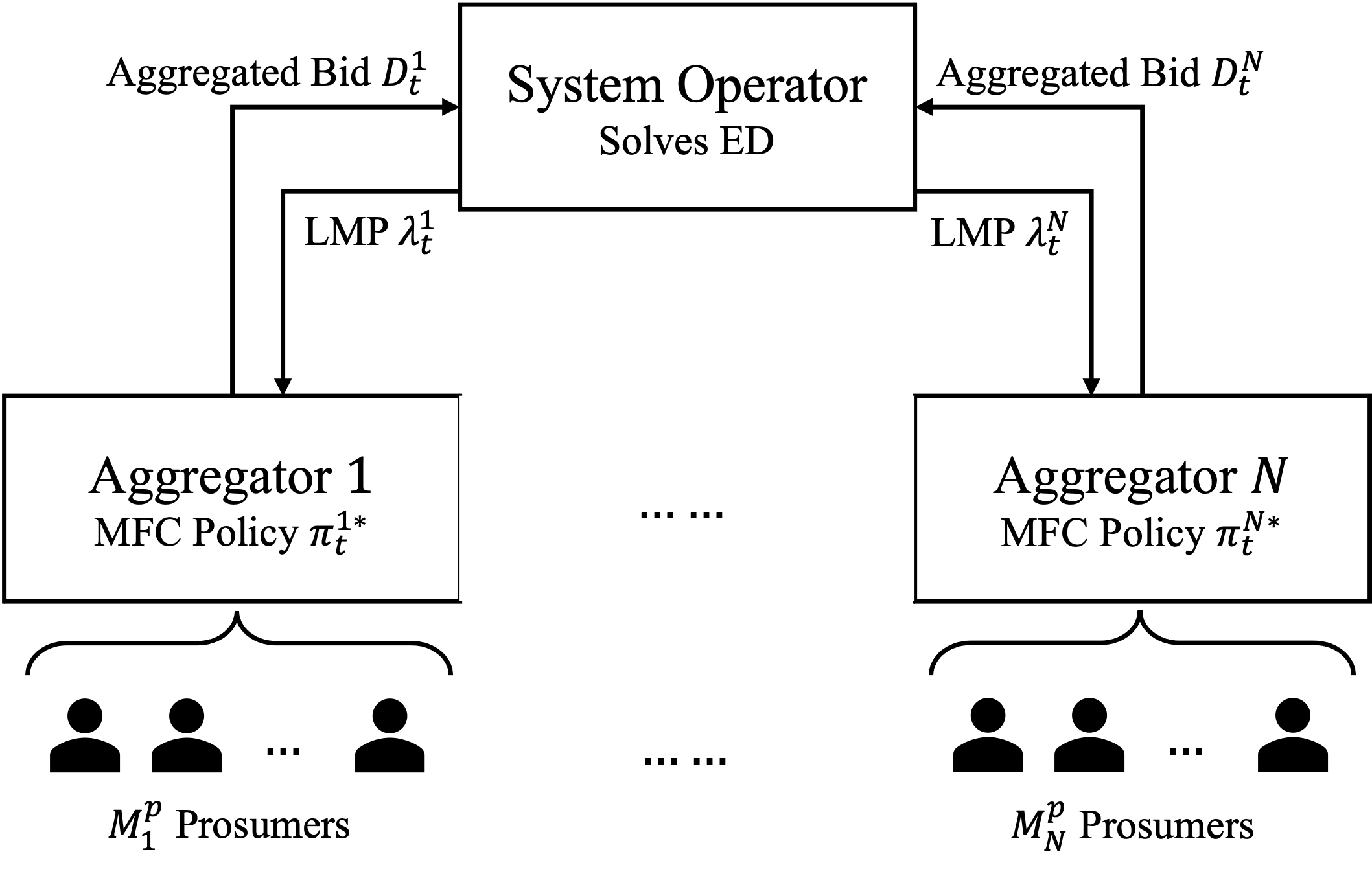}  
\caption{Illustration of the relationship between the single-aggregator MFC problem and the multi-aggregator MFE. The system operator clears the market by solving the ED problem, generating system-level LMPs. Each aggregator $n$ responds by solving its own MFC problem and adopting an optimal policy $\pi_t^{n*}$. An MFE is reached when the aggregate behavior induced by all aggregators' policies is consistent with the MF anticipated in their individual optimization problems.}
\label{fig:mfe-illus}
\end{figure}

Building on the MFC formulation at the individual aggregator level, in this section, we define an MFE among aggregators, which captures the steady-state outcome of their dynamic interplay. We then establish conditions under which the existence and uniqueness of such an MFE can be guaranteed. Figure~\ref{fig:mfe-illus} illustrates this transition from the single-aggregator MFC formulation to the multi-aggregator MFE setting: while each aggregator individually solves an MFC problem against LMPs determined by the system operator, the collection of aggregators, now indexed by $n = 1, \ldots, N$, jointly induces a MF that must be consistent with the policies they optimally adopt in equilibrium. The MF approximation is taken within each aggregator as the number of underlying prosumers grows large, yielding a MFC problem for each aggregator. Aggregators interact only through market clearing: their collective net-demand bids determine LMPs via ED, which couples the otherwise separate MFC problems through an endogenous price feedback loop. We use the terminology ``MFE'' to denote the fixed point between (i) MFC policies among finite number of aggregators and (ii) LMP consistency.

An MFE is a fixed point in a dynamic system of infinitely many interacting agents, characterized by two key properties: (i) \textbf{optimality}, where each agent’s strategy is optimal given the mean field, and (ii) \textbf{consistency}, where the mean field coincides with the distribution of states and actions induced by all agents following that strategy.

Note that an MFE is not generally a Nash equilibrium, as the latter is typically defined for games with a finite number of agents. Our equilibrium concept is a price-consistent equilibrium: each aggregator solves an MFC problem taking the LMP profile as given, and the LMP profile must be consistent with the ED outcome induced by all aggregators' net bids. This differs from a Nash equilibrium among aggregators, which would require each aggregator to explicitly account for how its actions affect prices through market clearing; we do not model such strategic price impact in the present work. Whether an MFE arises as the limit of $M$-agent Nash equilibria as $M\to \infty$ is a nontrivial question. \cite{Basar_MFE_Nash} provides sufficient conditions under which the infinite-agent MFE policy yields an $\eps$-Nash equilibrium for the corresponding finite-$M$ game. In this work, we focus on the properties of the MFE in the infinite-agent setting and leave its connection to finite-agent Nash equilibria to future research.

To define MFE formally, we introduce two operators: the optimality operator and the consistency operator. Let the space of state–dependent stochastic policies be denoted by $\Pi := \left\{ \pi \mid \pi: \mathcal{S} \times \RR^H \to \mathcal{P}(\mathcal{A}) \right\}$. The optimality operator is defined as follows.

\begin{definition}[Optimality Operator]
For each aggregator $n$, the optimality operator $\Gamma^n_1: \mathcal{S} \times \RR^H \to \Pi$ maps a state $s^n \in \mathcal{S}$ and LMP profile $\vlambda^n \in \RR^H$ to an optimal policy $\pi^{n*}$, i.e.,
\begin{equation*}
\pi^{n*} = \Gamma^n_1(s^n, \vlambda^n),
\end{equation*}
such that for any policy $\pi^n \in \Pi$,
\begin{equation*}
V^\reg_n(s^n, \pi^{n*}, \vlambda^n) \geq V^\reg_n(s^n, \pi^n, \vlambda^n).
\end{equation*}
Equivalently, the operator is defined as
\begin{equation*}
\Gamma^n_1(s^n, \vlambda^n) := \arg\max_{\pi^n \in \Pi} V^\reg_n(s^n, \pi^n, \vlambda^n).
\end{equation*}
\end{definition}
\noindent Note that the optimal policy $\pi^{n*}$ is not unique in general. However, our setup with strongly convex regularization term can ensure the uniqueness of optimal policy (see Proposition \ref{prop:reg-unique} in Appendix \ref{app:uniqueness}). Next, we define both the consistency operator and the concept of consistency: 
\begin{definition}[Consistency Operator and Consistency]
For each aggregator $n$, the consistency operator is a mapping
$
\Gamma^n_2: \mathcal{P}(\mathcal{S} \times \mathcal{A}) \times \Pi \to \mathcal{P}(\mathcal{S} \times \mathcal{A}),
$
which takes an MF $\mathcal{L}^n \in \mathcal{P}(\mathcal{S} \times \mathcal{A})$ and a policy $\pi^n \in \Pi$, and returns the updated MF  $ \mathcal{L}^{n\prime} $ resulting from all prosumers at bus $n$ following policy $\pi^n$.

A pair $(\mathcal{L}^n, \pi^n)$ satisfies the \emph{consistency condition} if the mean field is a fixed point of the operator; that is, $
\Gamma^n_2(\mathcal{L}^n, \pi^n) = \mathcal{L}^n.
$
\end{definition}
In our model, given the fixed LMP vector $\vlambda^n$, the consistency operator has the following form:
\begin{align}
    & \Gamma^n_2(\LL^n, \pi^n)(s', a') \coloneq \pi^n(a' \mid s', \vlambda^n)  \nonumber \\
    & \hspace{10pt} \cdot \left[\zeta U^n(s') + (1 - \zeta)\sum_{s, a}\LL^n(s, a) P^n(s' \mid s, a)\right]
    \label{eq:update-mf}
\end{align}
for all $s' \in \mathcal{S}, a' \in \mathcal{A}$, and $\zeta \in (0, 1)$ is the probability of a uniform noise with $U^n$ as the next-state distribution induced by the uniform storage reset. $P^n(s' \mid s, a')$ denotes the state transition probability for the aggregator at bus $n$. 

Let $\vpi = (\pi^{1}, \cdots, \pi^{N})$ denote the policy profile of all aggregators. We now give the formal definition of an MFE.
\begin{definition}[MFE]
A profile $\{(\pi^{n*}, \vlambda^{n*})\}_{n=1}^{N}$ is a mean field equilibrium if, for each aggregator $n$ and any state $s^n \in \mathcal{S}$ at time $t$, the following conditions hold:
\begin{enumerate}
    \item \textbf{Optimality:} Given $\vlambda^{n*}$, the optimal policy is obtained via the optimality operator: 
   $
    \pi^{n*} = \Gamma^n_1(s^n, \vlambda^{n*}).
  $
    \item \textbf{Consistency:} If all aggregators follow $\vpi^*$, then the mean field distribution evolves according to the consistency operator:
   $
    \LL^n_{t+1} = \Gamma^n_2(\LL^n_t, \pi^{n*}),
  $
    and the corresponding LMP obtained through the ED problem \eqref{eqn:ed-obj} -- \eqref{eqn:gen-limit}, 
    $\lambda^n_{t+1}(\vLL_{t+1})$, must satisfy the consistency condition $
    \lambda^n_{t+1}(\vLL_{t+1}) = \left[\vlambda^{n*}\right]_{\Thour(t+1)}.$
\end{enumerate}
\end{definition}

\noindent The definitions of the optimality and consistency operators, along with that of an MFE, naturally lead to a three-step computational framework:

\noindent\textbf{Step 1:} Fix $\LL^n_t$ and $\vlambda_t$. Given $s^n$, aggregator $n$ finds the optimal policy using the optimality operator: $\pi^{n*}_t = \Gamma^n_1(s^n, \vlambda_t)$.

\noindent\textbf{Step 2:} Update the mean field via the consistency operator: $\LL^n_{t+1} = \Gamma^n_2(\LL^n_t, \pi^{n*}_t)$. Then compute the updated LMP $\lambda^n_{t+1}(\vLL_{t+1})$ by solving the ED optimization problem \eqref{eqn:ed-obj} -- \eqref{eqn:gen-limit}. If $t+1$ marks the start of a new day, update $\vlambda^n$ with the most recent $H$ LMPs.

\noindent\textbf{Step 3:} Repeat Steps 1 and 2 until $\vlambda^n_{t+1}$ converges to $\vlambda_t$.

The key idea behind the computational framework is that information about the MF is embedded in the LMPs. At each timestep, aggregators update their policies based on observed LMPs, and the MF evolves in response to these updated policies. While this defines a structured framework, it is not a complete algorithm: Step 1 assumes access to an optimal policy, yet both the LMPs in the reward function~\eqref{eqn:rew} and the transition kernel $P^n(s' \mid s, a')$, which depends on unknown net load distributions, are not known to the aggregators. Section~\ref{sec:alg} addresses this challenge by introducing an RL algorithm to learn these unknown quantities. For now, assuming Step 1 can be executed, we show below that the 3-step procedure forms a contraction mapping, leading to a unique fixed point, which is exactly an MFE.

\subsection{Existence and Uniqueness of MFE under Regularization}
\label{sec:proof}
Our approach here is constructive: we prove that the proposed 3-step framework defines a contraction mapping over the space of policy and price profiles, guaranteeing convergence to a unique fixed point. This fixed point corresponds exactly to an MFE, thereby establishing its existence and uniqueness. Before presenting the main convergence result, we first show that both the optimality and consistency operators are Lipschitz continuous. We adopt the $\ell_1$-norm as policies and MF are probability measures over finite spaces, and the distance between 2 distributions can be measured using the total variation distance which is half of the $\ell_1$ distance. Proof sketches are provided in Appendix~\ref{app:pf-thm-op} and~\ref{app:pf-thm-co}.
 
\begin{theorem}[Lipschitz Continuity of the Optimal Solution Mapping $\Gamma^n_1$]
    Under the same assumptions as in Proposition \ref{prop:feng2025}, in the regularized MFG, for any belief $\vlambda^n, \vlambda^{n'}$, there exists $L_1 \geq 0$ such that:
    \begin{equation}
        \sup_{s \in \mathcal{S}} \normof{\Gamma^n_1(s, \vlambda^n) - \Gamma^n_1(s, \vlambda^{n'})}_1 \leq L_1 \normof{\vlambda^n - \vlambda^{n'}}_1.
        \label{eq:lip-gamma1}
    \end{equation}
    \label{thm:op}
\end{theorem}
\begin{theorem}[Lipschitz Continuity of $\Gamma^n_2$]
    Under the same assumptions as in Proposition \ref{prop:feng2025}, for any MF $\LL^n, \LL^{n'}$ and policies $\pi^n, \pi^{n'}$, given $\vlambda^n$, there exists $L_2, L_3 \geq 0$ such that
    \begin{align}
        & \normof{\Gamma^n_2(\LL^n, \pi^n) - \Gamma^n_2(\LL^{n'}, \pi^n)}_1 \leq L_2 \normof{\LL^n - \LL^{n'}}_1, \\
        & \normof{\Gamma^n_2(\LL^n, \pi^n) - \Gamma^n_2(\LL^n, \pi^{n'})}_1 \nonumber \\
        & \hspace{40pt} \leq L_3 \sup_{s \in \mathcal{S}} \normof{\pi^n(\cdot \mid s, \vlambda^n) - \pi^{n'}(\cdot \mid s, \vlambda^n)}_1.
    \end{align}
    \label{thm:gamma2}
\end{theorem}

We now present the main theorem regarding the MFE existence and uniqueness.
\begin{theorem}[Existence and Uniqueness of MFE]
    Assume the same assumptions as in Proposition \ref{prop:feng2025}, and assume that, for every admissible MF profile, the induced demand is feasible to ED. If $L_2+HN L_1L_3L_{\mathrm{MF}}<1$, the exact 3-step update converges to a unique MFE.
    \label{thm:mfe}
\end{theorem}
The specific Lipschitz constant $L_{\text{MF}}$ and the proof are provided in Appendix \ref{app:pf-mfe}.
\section{The Two-Phase RL Algorithm} 
\label{sec:alg}
The three-step procedure can be used directly when its exact optimality and consistency updates are available. Inspired by this procedure, we use finite-step RL training and hourly LMP-belief updates as a practical approximation. The hybrid MFC-MFG structure is not an arbitrary combination of methods: MFC models large DER portfolios, while market price clearing induces a fixed-point coupling across populations as a MFG. RL provides a computationally tractable mechanism to compute the resulting MFE when model parameters are partially unknown. We now turn to the key challenge of Step 1: computing the optimal policy $\pi^{n*}_t = \Gamma^n_1(s^n, \vlambda^n_t)$. To this end, we propose a two-phase, distributed mean-field RL algorithm executed within each time period $t$. Phase 1: Training -- each aggregator independently trains its policy using an RL algorithm for a specified number of steps, based on either LMP forecasts or prior beliefs about the LMPs for time period $t$.
Phase 2: Execution -- all prosumers act based on their aggregator’s trained policy. Since the policies are stochastic, even identical prosumers may take different actions. These actions are then compiled by the aggregator  into bids submitted to the market clearing process. LMPs for time $t$ are then calculated by solving the ED problem.  

Aggregators still do not have direct knowledge of external uncertainty distributions, such as those governing net load. However, a key feature of electricity markets can be leveraged: fluctuations in LMPs implicitly reflect both market dynamics and external uncertainties. Using a similar approach as in~\cite{feng2025decentralized}, we allow each aggregator to form and maintain its own belief about the LMPs for every timestep of the day, represented as a vector of length $H$. These beliefs, derived from historical observations, guide the aggregator’s RL-based decision-making in Phase 1. After the LMPs are realized through market clearing, aggregators update their beliefs adaptively.

We assume that all prosumers at the same bus share the LMP belief vector maintained by their aggregator, as the aggregator represents their collective interests and the actual LMP is identical for all agents at the same bus. Let $\vbelief^n_t \in \mathbb{R}^H$ denote the belief vector maintained by aggregator $n$ at time $t$. Once the system operator solves the ED problem and returns the realized LMP $\lambda^n_t$ for bus $n$, the aggregator updates its belief according to:
\begin{equation}
    \vbelief^n_{t+1} := \vbelief^n_t - \frac{\delta_n}{\Tday(t) + 1} \left((\vbelief^n_t)^\top \mathbf{1}_{\Thour(t)} - \lambda^n_t\right) \mathbf{1}_{\Thour(t)}
    \label{eqn:lmp-belief}
\end{equation}
where $\delta_n \in [0.5, 1]$ is a learning rate hyper-parameter for this LMP update rule, and $\mathbf{1}_{\Thour(t)} \in \RR^H$ denotes the vector whose value is 1 at the $\Thour(t)$-th entry and 0 everywhere else. That is, the update rule changes only the $\Thour(t)$-th entry of the belief vector at each time $t$. 

\textbf{Training Phase:} At the beginning of each time period $t$, each aggregator fixes its LMP belief and trains a policy using an RL algorithm, referred to generically as \textit{Alg} for $\Ttrain$ steps.

\textbf{Execution Phase:} After each aggregator has learned a policy, this policy is distributed to its prosumers. At bus $n$, prosumer $i$ takes an action from the policy $a^n_{it} \sim \pi^{n*}_t(\cdot | s^n_{it}, \vbelief^n_t)$. Also, each prosumer $i$ and consumer $j$'s original net demand are realized following the distributions $\mathcal{Q}^p_n$ and $\mathcal{Q}^c_n$, respectively. The net demand in the quantity of energy now has the following forms for prosumer $i$ at bus $n$ at time $t$:
\begin{equation}
    d^n_{it} = (\Phi(e^n_{t}, a^n_{t}, \eta_n) + \nd^n_{t}) \overline{E}^n_i,
\end{equation}
and similarly, for each consumer $j$ at bus $n$ at time $t$:
\begin{equation}
    d^n_{jt} = \nd^n_{jt} \overline{E}^n_j.
\end{equation}

Recall that we introduce a random regeneration probability $\zeta$, which gives each prosumer a positive probability of changing its storage capacity. This serves two purposes: it maintains a dynamic environment to encourage continual learning, even at a steady state, and it mimics real-world conditions where prosumers may enter or exit the system at a given location. The transition of the state of charge for each prosumer $i$ at bus $n$ from time $t$ to $t+1$ is defined as follows: 
\begin{equation}
    e^n_{i,t+1} = \begin{cases}
        \text{Uniform}(0, 1), \text{with probability $\zeta$}, \\
        \max \left\{\min \left\{e^n_{it} + a^n_{it}, 1\right\}, 0 \right\}, \text{with probability $1-\zeta$}.
    \end{cases}
    \label{eqn:random-regen}
\end{equation}
Consequently, the aggregator at $n$ updates its storage level as:
\begin{equation}
    e^n_{t+1} = \frac{1}{\overline{E}_n}\sum_{i=1}^{M^p_n} e^n_{i,t+1} \overline{E}^n_i,
    \label{eqn:agg-soc-transition}
\end{equation}
which is a weighted average of all its prosumers storage level. This storage level is then used as the initial storage level state to begin the RL training for time $t$ for $\Ttrain$ steps. With this setup, we present the pseudo-code in Algorithm \ref{alg:two-phase}.

\begin{algorithm}
    \caption{A two-phase distributed MF-RL algorithm with LMP beliefs and entropy regularization}
    \label{alg:two-phase}
    \KwIn{Initial battery states $e^n_0 \in [0, 1]$, initial LMP beliefs $\vbelief^n_0 \in \mathbb{R}^H$ with learning rates $\delta_n \in [0.5, 1]$, demand shapes $\mathcal{Q}^p_n, \mathcal{Q}^c_n$; training step $\Ttrain$; random regeneration probability $\zeta$; an RL algorithm \textit{Alg}; time functions $\Thour(\cdot), \Tday(\cdot)$.}
    \For{$t = 0, 1, \ldots$}{
        \vspace{5pt}
        \textbf{\textit{Training phase}} \\
        \For{Bus $n = 1, \ldots, N$}{
            Train the aggregator for $\Ttrain$ steps using \textit{Alg} with initial storage $e^n_t$ under $\vbelief^n_t$ to get the optimal policy $\pi^{n*}_t$ \;
        }
        \vspace{5pt}
        \textbf{\textit{Execution phase}} \\
        \For{Bus $n = 1, \ldots, N$}{
            \For{Prosumer $i = 1, \ldots, M^p_n$}{
                Get net demand $\nd^n_{it} \sim \mathcal{Q}^p_n$ \;
                Take actions $a^n_{it} \sim \pi^{n*}_t$ \;
                Storage state transition as in \eqref{eqn:random-regen} \;
            }
            \For{Consumer $j = 1, \ldots, M^c_n$}{
                Get demand $\nd^n_{jt} \sim \mathcal{Q}^c_n$ \;
            }
            Compute next storage state $e^n_{t+1}$ as in \eqref{eqn:agg-soc-transition} \; 
        }
        Solve an ED problem to get LMPs $\lambda^n_t$ for all $n$ \;
        Update the LMP belief as in \eqref{eqn:lmp-belief} \;
    }
\end{algorithm}

\textit{Remark on distribution network constraints.} 
The baseline formulation models aggregators at transmission-level buses and therefore does not explicitly include distribution network constraints. However, such constraints can be incorporated into our framework without modifying its overall structure. In particular, voltage or line-flow limits at the distribution level can be enforced through negative reward terms during RL training. This approach follows the same principle used in peer-to-peer energy trading in distribution networks \cite{P2P_FengLiu}. Such extensions are not pursued in this work, as our focus is on presenting the framework for wholesale-level market interactions. 
\section{Numerical Experiment} \label{sec:numerical}
In this section, we conduct numerical experiments to analyze market outcomes under the proposed framework.

\subsection{Test Case}
We use the 37-bus synthetic Oahu, Hawaii network from \cite{birchfield}, which provides bus coordinates. To align with real-world data, we map each power plant operated by Hawaiian Electric, the sole utility provider on Oahu, to its nearest bus in the synthetic network, using data from \cite{powerfacts}. The modified system has 26 generators: 4 oil, 2 biomass, 17 utility-scale solar (distinct from household PV), and 3 wind. Oil and biomass units follow quadratic costs $C(p)=ap^2+bp$, with coefficients adapted from \cite{isone, tidball2010cost}: oil $a \in [0.0059,\,0.0342]$, $b=19.98$; biomass $a \in [0.001,\,0.002]$, $b \in [28.45,\,52.65]$. Specific generators' coefficients are uniformly sampled from these ranges.

The generation cost for solar and wind units is set to zero. Their capacity factors are treated as random variables. To reflect temporal patterns without explicitly modeling time series dependencies, we use historical hourly capacity factor profiles as the expected values, with utility-scale solar data from \cite{profilesolar} and wind from \cite{argueso}. At each timestep, the average capacity factor is multiplied by an independent random noise factor: a triangular distribution \( \Delta(0.8, 1.2, 1) \) for solar and \( \Delta(0.5, 1.5, 1) \) for wind. Here, \( \Delta(\alpha, \beta, m) \) denotes a triangular distribution with lower limit \( \alpha \), upper limit \( \beta \), and mode \( m \).

Each bus in the network hosts two  650 prosumers and 2,000 pure consumers. Prosumers are further categorized into three types, 500 small, 100 medium, and 50 large, equipped with storage capacities of 10, 20, and 30~kWh, respectively. 
Daily demand profiles are based on hourly average data from \cite{coffman2016estimating}. For prosumers, demand captures net load (i.e., after behind-the-meter DER output), while for pure consumers, it reflects gross load. The dataset provides a single standardized net load profile (percentage of capacity) for all prosumers. Thus, larger storage capacities imply proportionally greater solar generation and consumption.
To introduce demand uncertainty, we scale each agent's average daily demand by a factor drawn from the triangular distribution \( \Delta(0.8, 1.2, 1) \). 

\subsection{Model Predictive Control (MPC) Benchmark}
\label{subsec:mpc}
We compare the proposed MFE learning framework against a centralized forecast-based MPC benchmark. The MPC serves as a standard optimization-based baseline commonly used in power system operations. At each timestep, the MPC solves a finite-horizon optimal power flow problem with a lookahead window of length $H$ (one day), which consists of 1 deterministic current stage followed by $H-1$ future stages. Uncertainty in demand and renewable generation is handled using the same belief vectors as in~\eqref{eqn:lmp-belief}. 

We keep the notation of variables as in Sections~\ref{sec:Wholesale} and~\ref{sec:Agg} for the current stage at time $t$. For the future stages $h\in \mathcal{H} := \{1,\dots,H-1\}$, let $\widehat{\nd}^n_{i,t+h}$ and $\widehat{\nd}^n_{j,t+h}$ denote the (exogenous) fixed demand belief at bus $n$ for prosumer $i$ and consumer $j$, respectively; and let $\widehat{p}_{g,t+h}$ denote the capacity factor belief for renewable generator $g$. Because now all batteries' charging and discharging decisions are being made through a central controller, we introduce the decision variables $a^{\mathrm{C}}_{nt}$ and $a^{\mathrm{D}}_{nt}$ as the charging and discharging variables at current stage, and $e^n$ as the battery level at the end of current stage. For each future stage at $t+h$, we analogously define $a^{\mathrm{C}}_{n,t+h}$, $a^{\mathrm{D}}_{n,t+h}$ and $e^n_{t+h}$ for charging, discharging and battery level, respectively. The net demand as defined in~\eqref{eq:Demand_n} is now rewritten as:
\begin{align}
    & D^n_t = \sum_{i=1}^{M^p_n} \nd^n_{it} + \sum_{j=1}^{M^c_n} \nd^n_{jt} + \overline{E}^n \cdot \left(\eta_n a^{\mathrm{C}}_{nt} - \frac{1}{\eta_n} a^{\mathrm{D}}_{nt} \right), \forall n \in \mathcal{N}, \label{eq:Demand_n_curr} \\
    & D^n_{t+h} = \sum_{i=1}^{M^p_n} \widehat{\nd}^n_{i,t+h} + \sum_{j=1}^{M^c_n} \widehat{\nd}^n_{j,t+h} \nonumber \\
    & \hspace{25pt} +  \overline{E}^n \left( \eta_n a^{\mathrm{C}}_{n,t+h} - \frac{a^{\mathrm{D}}_{n,t+h}}{\eta_n} \right), \forall n \in \mathcal{N}, h \in \mathcal{H}. \label{eq:Demand_n_future}
\end{align}
where $\eta_n$ is the one-way battery efficiency for aggregator $n$. For each future stage, the power balance, transmission limit and generator capacity constraints are defined as:
\begin{align}
    & \sum_{g=1}^G p_{g,t+h} = \sum_{n=1}^{N} \frac{D^n_{t+h}}{\Delta t}, \; \forall h \in \mathcal{H}, \label{eqn:power-balance-future} \\
    & -\overline{F}_l \leq \sum_{n=1}^{N} \text{PTDF}_{ln} \left( \sum_{g \in G_n} p_{g,t+h} - \frac{D^n_{t+h}}{\Delta t} \right) \leq \overline{F}_l, \nonumber \\
    & \hspace{150pt} \forall l \in \mathcal{L}, h \in \mathcal{H}, \label{eqn:flow-limit-future} \\
    & 0 \leq p_{g,t+h} \leq \widehat{p}_{g,t+h}, \; \forall g \in \mathcal{G}. \label{eqn:gen-limit-future}
\end{align}
In addition, the battery level constraints are:
\begin{align}
    & e^n_t = e^n_{t-1} + \eta_n a^{\mathrm{C}}_{nt} - \frac{a^{\mathrm{D}}_{nt}}{\eta_n}, \; \forall n \in \mathcal{N}, \label{eq:soc_n_curr} \\
    & e^n_{t+h} = e^n_{t+h-1} + \eta_n a^{\mathrm{C}}_{n,t+h} - \frac{a^{\mathrm{D}}_{n,t+h}}{\eta_n}, \forall n \in \mathcal{N}, h \in \mathcal{H}, \label{eq:soc_n_future} \\
    & 0 \leq e^n_t, \eta_n a^{\mathrm{C}}_{nt}, \frac{a^{\mathrm{D}}_{nt}}{\eta_n} \leq 1, \forall n \in \mathcal{N} \label{eq:current_soc_bounds} \\
    & 0 \leq e^n_{t+h}, \eta_n a^{\mathrm{C}}_{n,t+h}, \frac{a^{\mathrm{D}}_{n,t+h}}{\eta_n} \leq 1, \forall n \in \mathcal{N}, h \in \mathcal{H} \label{eq:future_soc_bounds}
\end{align}
where $e^n_{t-1}$ is the battery level at the end of $t-1$, and is an input already computed at $t-1$. We now present the complete \textit{MPC optimization problem} at time $t$ as follows:
\begin{align}
    & \minimize \sum_{g\in\mathcal{G}} \left( C_g(p_{gt}) + \sum_{h \in \mathcal{H}} C_g(p_{g,t+h}) \right) \\
    & \text{s.t.} ~\eqref{eqn:power-balance}-\eqref{eqn:gen-limit}, \text{and}~\eqref{eq:Demand_n_curr}-\eqref{eq:future_soc_bounds}.\nonumber
\end{align}
After solving the optimization problem, the MPC commits only the current-stage storage actions $\{a^{\mathrm{C}}_{nt}, a^{\mathrm{D}}_{nt}\}_{n\in\mathcal{N}}$, updates the battery storage, proceed to the next timestep and repeats. Actual fuel costs, generation, transmission and LMPs are recovered from the dual variables by resolving the one-stage ED problem in Section~\ref{sec:Wholesale}. The MPC benchmark represents a centralized controller with full system information and optimization using forecasts of future demand and renewable availability. In contrast to the proposed MFE approach, MPC does not rely on decentralized learning or equilibrium convergence, but instead repeatedly solves a large-scale optimization problem at each timestep.

\subsection{Decentralized Heuristic Algorithm (DHA) Benchmark}
\label{subsec:ha}
We also compare the proposed MFE learning framework against a decentralized heuristic algorithm (DHA), which serves as a simple rule-based baseline. At each timestep $t$, each aggregator $n$ determines a storage action by first using the same LMP belief vector as in~\eqref{eqn:lmp-belief}. Then each aggregator $n$ defines the low- and high-price windows using pre-determined and fixed thresholds $\lambda_{\mathrm{low}}^n$ and $\lambda_{\mathrm{high}}^n$ (where $\lambda_{\mathrm{high}}^n>\lambda_{\mathrm{low}}^n$) over the next day:
\begin{align}
\mathcal{T}^n_{\mathrm{low}}(t) := \left\{ h: \left[\vbelief^n_t\right]_h \le \lambda_{\mathrm{low}}^n \right\}, \label{eq:ha_low_window}\\
\mathcal{T}^n_{\mathrm{high}}(t) := \left\{ h: \left[\vbelief^n_t\right]_h \ge \lambda_{\mathrm{high}}^n \right\} \label{eq:ha_high_window},
\end{align}
where $\left[\vbelief^n_t\right]_h$ indicates the $h$-th entry of the vector $\vbelief^n_t$. The cardinality of the two sets can be computed as $N^n_{\mathrm{low}}(t):=|\mathcal{T}^n_{\mathrm{low}}(t)|$ and $N^n_{\mathrm{high}}(t):=|\mathcal{T}^n_{\mathrm{high}}(t)|$.

DHA spreads charging across the low-price window rather than charging aggressively at a single hour. If $N^n_{\mathrm{low}}(t)>0$, define the target battery level at the end of the low-price window as
\begin{equation}
e^{n*}_{\mathrm{ch}, t} := \min \left\{1, e^n_t + N^n_{\mathrm{low}}(t)(1-e^n_t)\right\}, \label{eq:ha_target_charge}
\end{equation}
and define the planned per-step charging amount as
\begin{equation}
\overline{a}^C_{nt} := \frac{e^{n*}_{\mathrm{ch}, t}-e^n_t}{N^n_{\mathrm{low}}(t)}. \label{eq:ha_plan_charge}
\end{equation}
Similarly, if $N^n_{\mathrm{high}}(t)>0$, define the target battery level at the end of the high-price window as
\begin{equation}
e^{n*}_{\mathrm{dis},t} := \max\{0,\ e^n_t - N^n_{\mathrm{high}}(t)\,e^n_t\},
\label{eq:ha_target_dis}
\end{equation}
and define the planned per-step discharging amount as
\begin{equation}
\overline{a}^{D}_{nt} := \frac{e^n_t-e^{n*}_{\mathrm{dis},t}}{N^n_{\mathrm{high}}(t)}. \label{eq:ha_plan_dis}
\end{equation}
Now, given the current level $e^n_t$, the deterministic action $a^{n*}_t\in[-1,1]$ is
\begin{equation}
a^{n*}_t=
\begin{cases}
-\min\{\overline{a}^{D}_{nt},\, e^n_t\}, &\text{if } T_{\mathrm{hour}}(t) \in \mathcal{T}^n_{\mathrm{high}}(t),\\
\min\{\overline{a}^C_{nt},\, 1-e^n_t\}, &\text{if } T_{\mathrm{hour}}(t) \in \mathcal{T}^n_{\mathrm{low}}(t),\\
0, &\text{otherwise},
\end{cases}
\label{eq:ha_det_action}
\end{equation}
where we prioritize discharging if both windows are active. To reduce synchronized behavior, each aggregator draws from an i.i.d. random factor $\xi^n_t \sim \mathrm{Uniform}(\alpha,1)$, where $\alpha\in(0,1)$, and implements the final action
\begin{equation}
a^n_t := \xi^n_t\, a^{n*}_t. \label{eq:ha_action}
\end{equation}
Finally, the battery level is updated via \eqref{eqn:storage-transition} and the resulting aggregate net demand bids are cleared by the standard ED \eqref{eqn:ed-obj}--\eqref{eqn:gen-limit}. The algorithm pseudo-code is presented in Algorithm~\ref{alg:dha}.

\begin{algorithm}
    \caption{A Decentralized Heuristic Algorithm}
    \label{alg:dha}
    \KwIn{Initial battery states $e^n_0 \in [0, 1]$, initial LMP beliefs $\vbelief^n_0 \in \mathbb{R}^H$ with learning rates $\delta_n \in [0.5, 1]$, demand shapes $\mathcal{Q}^p_n, \mathcal{Q}^c_n$; thresholds $\{\lambda_{\mathrm{low}}^n,\lambda_{\mathrm{high}}^n\}_{n\in\mathcal{N}}$; randomization parameter $\alpha\in(0,1)$, time functions $\Thour(\cdot), \Tday(\cdot)$.}
    \For{$t = 0, 1, \ldots$}{
        \vspace{5pt}
        \textbf{\textit{Training phase}} \\
        \For{Bus $n = 1, \ldots, N$}{
            Compute $\mathcal{T}^n_{\mathrm{low}}(t), \mathcal{T}^n_{\mathrm{high}}(t)$ via~\eqref{eq:ha_low_window}-~\eqref{eq:ha_high_window}\;
            Compute $\overline{a}^C_{nt}$ and $\overline{a}^{D}_{nt}$ via \eqref{eq:ha_plan_charge}--\eqref{eq:ha_plan_dis}\;
            Set $a^{n*}_t$ via \eqref{eq:ha_det_action} \;
            Draw $\xi^n_t\sim\mathrm{Uniform}(\alpha,1)$ and set $a^n_t=\xi^n_t a^{n*}_t$.
        }
        Submit the implied net demand bids and solve an ED problem to get LMPs $\lambda^n_t$ for all $n$ \;
        Update the LMP belief as in \eqref{eqn:lmp-belief} \;
        Update $\{e^n_{t+1}\}$ using \eqref{eqn:storage-transition}.
    }
\end{algorithm}

\subsection{Results and Discussion}
\begin{figure}[!ht]
    \begin{center}
        \begin{tabular}{@{}c@{}}
          \centering
          \includegraphics[width=0.85\columnwidth]{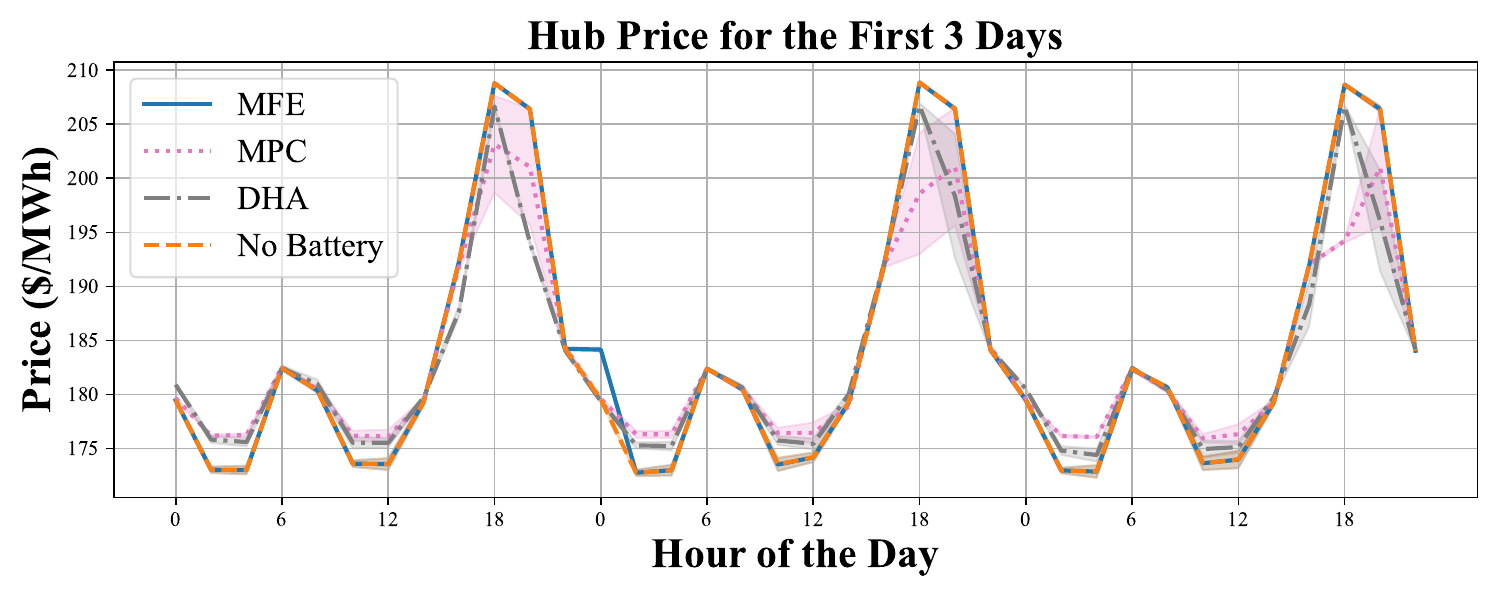}  
          \label{fig:hub-prices-first}
        \end{tabular}
        \begin{tabular}{@{}c@{}}
          \centering
          \includegraphics[width=0.85\columnwidth]{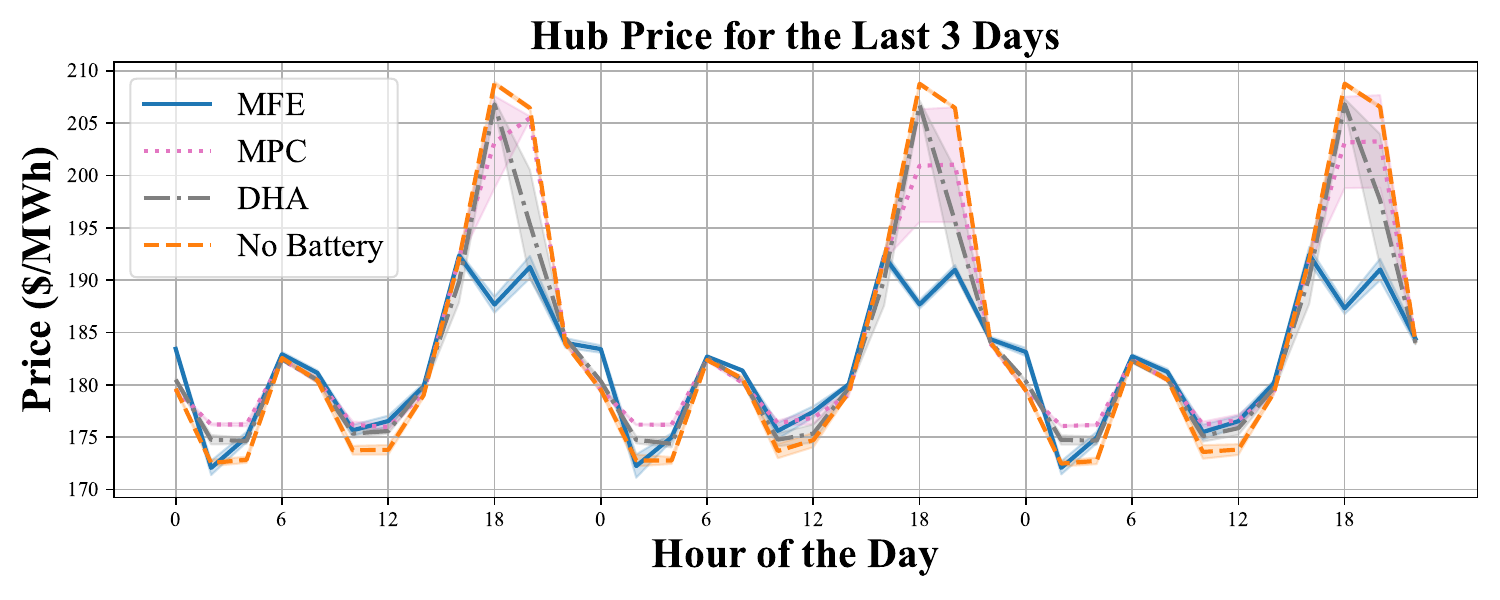}  
          \label{fig:hub-prices-last}
        \end{tabular}
        \caption{Comparison of hub prices among the 3 algorithms and the scenario without battery storage over the first 3 days (top) and last 3 days (bottom) during training. Shaded areas show the one standard deviation error bounds across all simulations.}
        \label{fig:hub-prices}
    \end{center}
\end{figure}

We employ proximal policy optimization (PPO) as the learning algorithm and set $T_{\text{train}} = 1,200$, with $H=12$ (corresponding to 2-hour timesteps), and $\delta_n = 0.9$ for all $n$, $\zeta = 0.01$. For benchmark, we set the same $\delta_n = 0.9$ for all belief's learning rates in MPC, and $\lambda_{\mathrm{low}}^n=180, \lambda_{\mathrm{high}}^n=190$ for all $n$, with $\alpha = 0.8$ for DHA. All simulation instances are run for a period of 50 days and repeated 10 times using different random seeds. The experiments were conducted on a Windows 11 system equipped with  Gen Intel(R) Core(TM) i7-13700KF (24 cores) and NVIDIA GeForce RTX 4070. The 2-hour timestep used in our experiments is a modeling choice intended to reduce state dimensionality and highlight intertemporal coordination effects. The algorithm itself is not tied to this discretization and can be implemented at finer resolutions (\textit{e.g.}, 15 minutes) without modification to the theoretical framework. Importantly, policy evaluation during execution is separate from RL training. We do not assess whether the training time meets a market-clearing deadline. Real-time operation requires only evaluating the trained policy and does not introduce additional computational cost at market clearing timescales. Figure \ref{fig:hub-prices} show the hub prices over the first and last 3 days of the simulation. Each plot compares the MFE with MPC and DHA, and if needed, also the scenario without storage (that is, prosumers only have grid-tied solar PVs).

\begin{figure}[!ht]
    \centering
    \includegraphics[width=0.85\columnwidth]{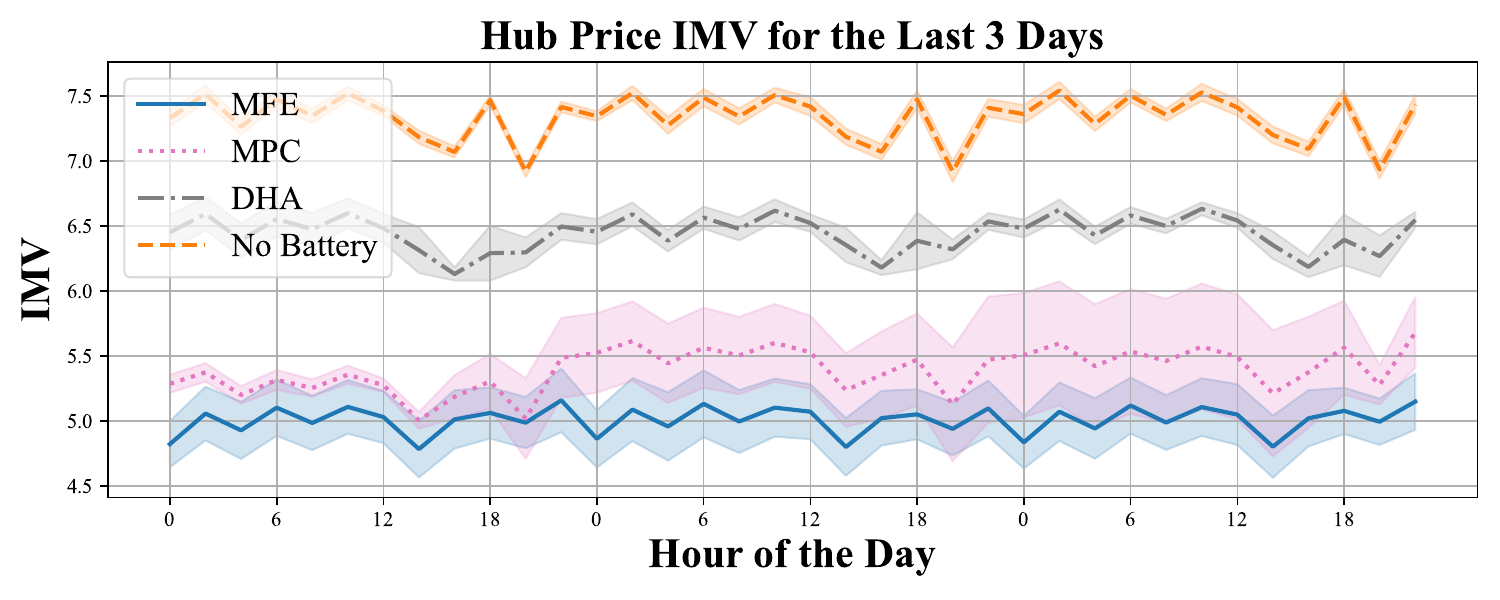}  
    \caption{Comparison of IMV of the last 3 days among the 3 algorithms and the scenario without battery storage. Shadow areas show the one standard deviation error bounds across all simulations.}
    \label{fig:imv}
\end{figure}

As shown in Figure \ref{fig:hub-prices}, at the start of the training and learning process, LMP hub prices behave similarly under all 4 scenarios. As learning progresses, prices under MFE become the least volatile. Additionally, the price curve with storage stabilizes and develops a consistent daily pattern, suggesting that a steady state has been reached.
To better measure the impacts of storage and RL on price volatility, we adopt the incremental mean volatility (IMV) measure from \cite{roozbehani}, which is defined as:
\begin{equation}
  \mathrm{IMV} = \lim_{T \to \infty} \frac{1}{T} \sum_{t=1}^T \abs{\lambda_{t+1} - \lambda_{t}}.
  \label{eqn:measure}
\end{equation}
For observed prices over a fixed periods from $t_0$ to $t_1$, we define the finite-window IMV as
\begin{equation}
  \mathrm{IMV}_{t_0:t_1} = \frac{1}{t_1 - t_0} \sum_{t=t_0}^{t_1-1} \abs{\lambda_{t+1} - \lambda_{t}}.
  \label{eqn:measure-finite-window}
\end{equation}
Figure~\ref{fig:imv} presents the IMV results over a rolling finite-window of 3 days among the 4 scenarios for the last 3 days. Both MPC and MFE scenarios yield significantly lower IMVs, confirming reduced price volatility. We also note that the MFE, as a decentralized learning algorithm which reacts based on price signals, has achieved the lowest IMV.

\begin{figure}[!ht]
    \begin{center}
        \begin{tabular}{@{}c@{}}
          \centering
          \includegraphics[width=0.85\columnwidth]{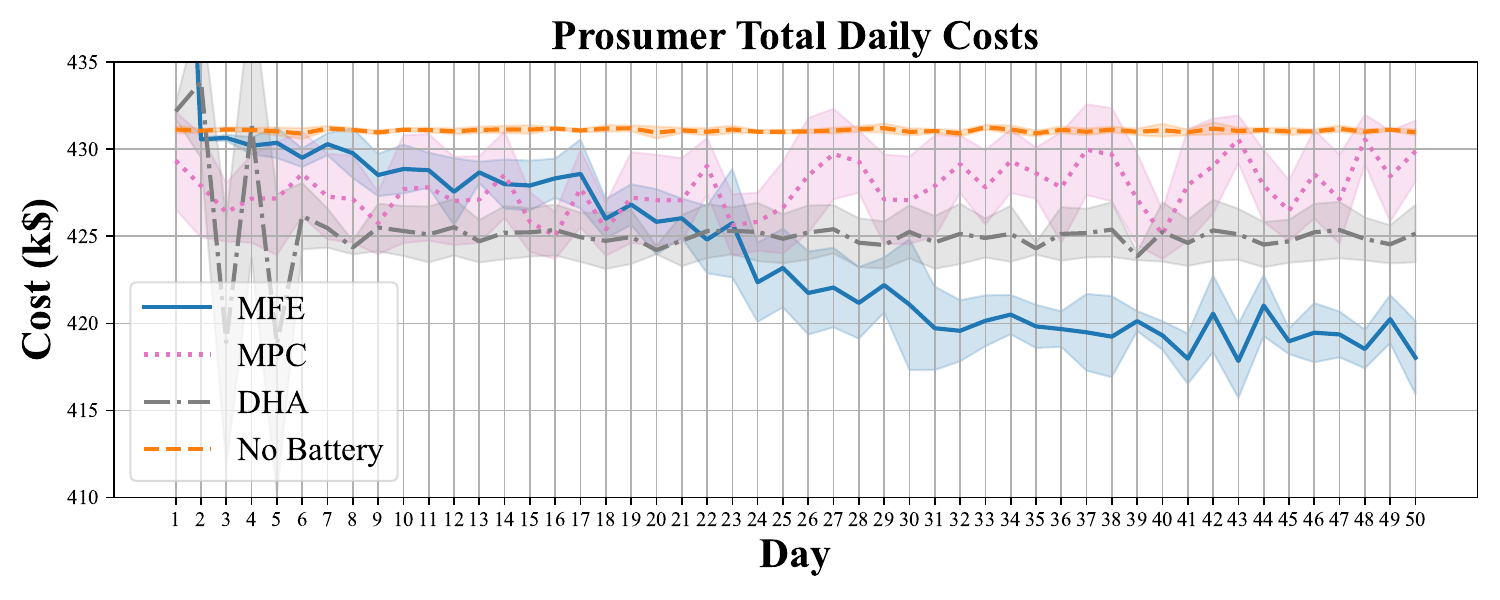}  
          \label{fig:cost-prosumer}
        \end{tabular}
        \begin{tabular}{@{}c@{}}
          \centering
          \includegraphics[width=0.85\columnwidth]{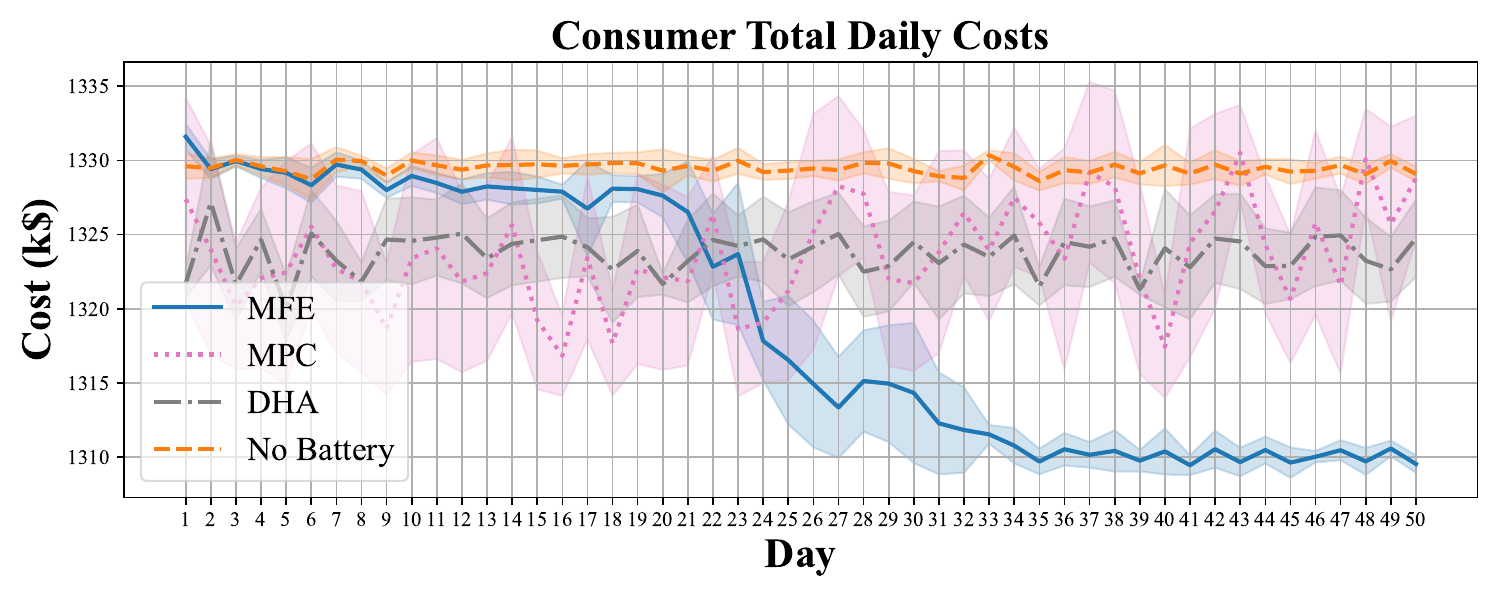}  
          \label{fig:cost-consumer}
        \end{tabular}
        \caption{Comparison of prosumers and consumers' daily costs, calculated as the sum of each time's LMP times the energy demand. Shaded areas show the one standard deviation error bounds across all simulations.}
        \label{fig:cost}
    \end{center}
\end{figure}

We also compute the average \textit{ex-post} daily cost, defined as the product of bidding quantity in MWh and real-time LMP, for all prosumers and consumers in the system. From Figure~\ref{fig:cost}, we observe that system-wide total daily costs are the lowest in the MFE scenario for both prosumers and consumers while the MPC and DHA both give slightly lower costs compared to the no-battery case. This demonstrates that our MFE framework delivers tangible cost benefits to all customers, not just prosumers.

\begin{figure}[!ht]
    \begin{center}
        \begin{tabular}{@{}c@{}}
          \centering
          \includegraphics[width=0.85\columnwidth]{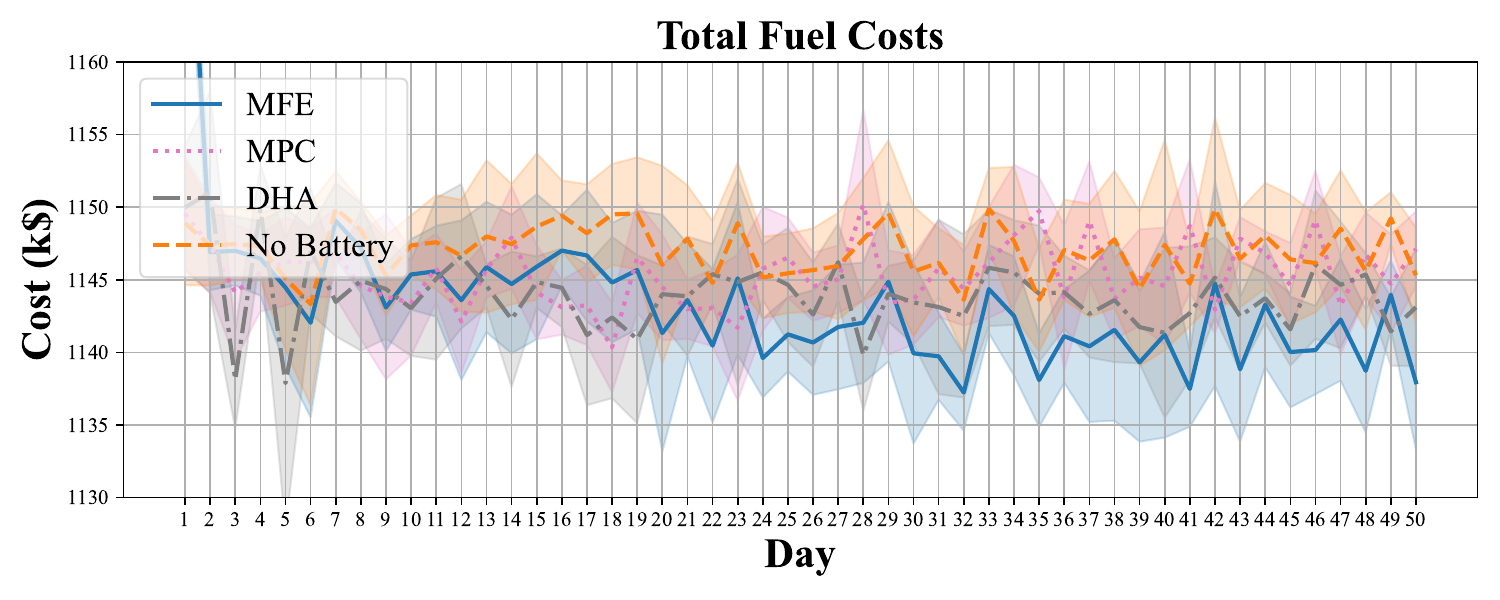}  
          \label{fig:obj-fuel}
        \end{tabular}
        \begin{tabular}{@{}c@{}}
          \centering
          \includegraphics[width=0.85\columnwidth]{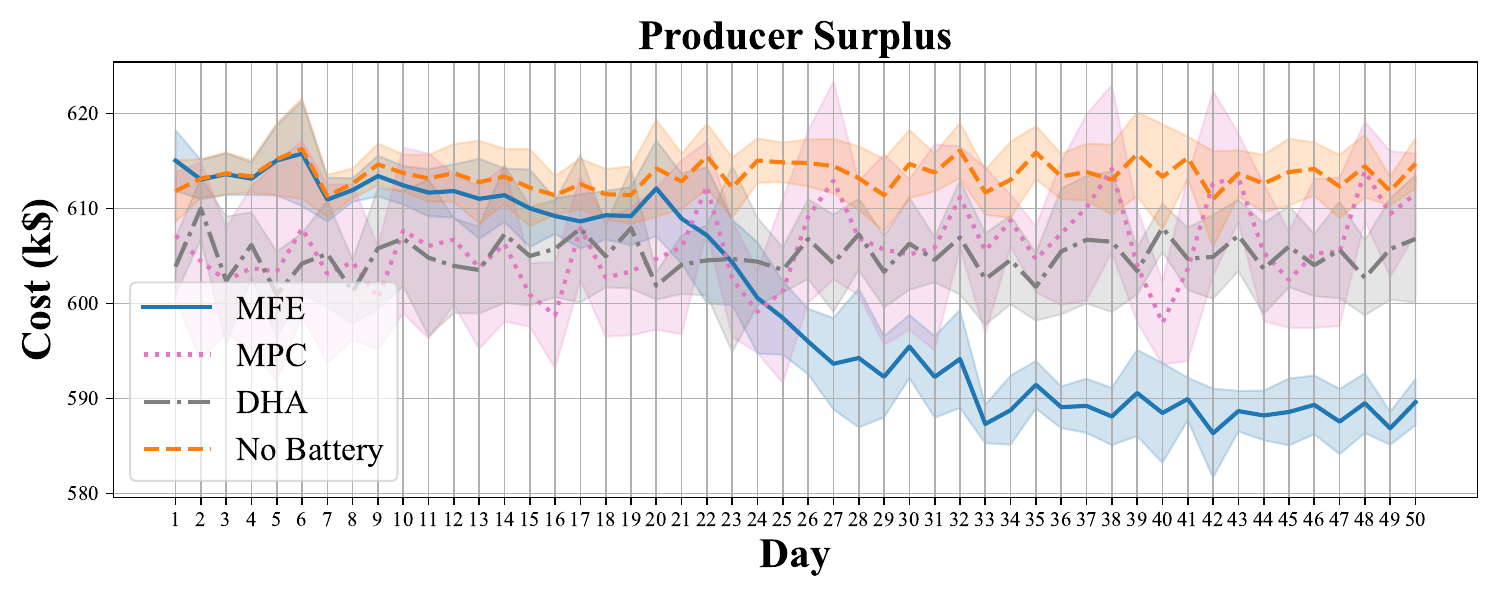}  
          \label{fig:producer-surplus}
        \end{tabular}
        \caption{Comparison of realized daily fuel cost (top), defined as the objective value of the one-stage economic dispatch problem, and daily producer surplus (bottom), defined as the difference between total customer payments and total fuel cost. Shaded areas indicate one standard deviation error bounds across all simulations.}
        \label{fig:fuel-ps}
    \end{center}
\end{figure}

We next examine system-level economic outcomes by comparing realized daily fuel cost and producer surplus, defined as total customer payments minus total fuel cost.\footnote{The simulation has no transmission congestion. By power balance, total customer payments equal total generator revenue. Customer payments minus fuel cost consequently equals generator revenue minus fuel cost, which is the producer surplus.} As shown in Figure~\ref{fig:fuel-ps}, the MFE scenario consistently achieves the lowest daily fuel cost over the 50-day horizon, followed by DHA and MPC, with the no-battery case being the most expensive. A similar ordering is observed in the producer surplus plot: a lower surplus implies that the system operates in a more economical way. Taken together, these results suggest that the MFE framework not only reduces price volatility and customer payments, but also improves allocative efficiency at the system level.

The MFE-based method exhibits the lowest realized fuel cost and producer surplus in our simulations, along with smoother aggregate demand and lower price volatility. Its policy updates use observed LMPs through the belief mechanism. The MPC benchmark solves a centralized finite-horizon problem using point forecasts and accounts for storage decisions in its dispatch optimization. DHA applies fixed thresholds to updated price beliefs, so its actions can also change as those beliefs change. 

\begin{figure}[!ht]
    \centering
    \includegraphics[width=0.85\columnwidth]{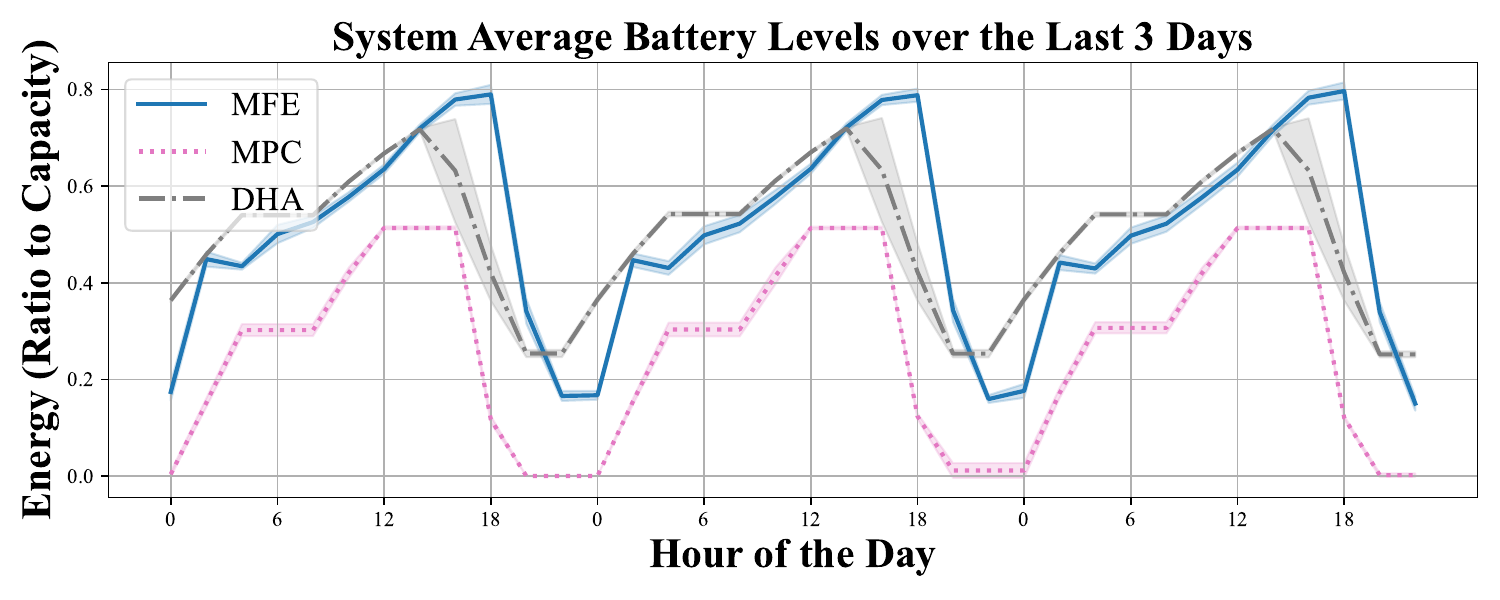}  
    \caption{Average battery level over the last 3 days across 10 simulations: the lines represent the system-average battery level compared among the 3 algorithms. 
    Shaded areas and error bars show the one standard deviation error bounds across all simulations.}
    \label{fig:state-action}
\end{figure}

We now observe the system-wide average battery level of prosumers. Figure~\ref{fig:state-action} shows the result over the final 3 days of simulation, averaged across 10 runs.
A clear daily pattern emerges from all 3 scenarios: prosumers tend to charge their storage during late-night and midday hours and discharge during morning and evening hours. We note that with our MFE approach, the battery level converges to a consistent daily cycle, indicating stabilization in agents' behavior and state evolution.

\begin{figure}[!ht]
    \centering
    \includegraphics[width=0.85\columnwidth]{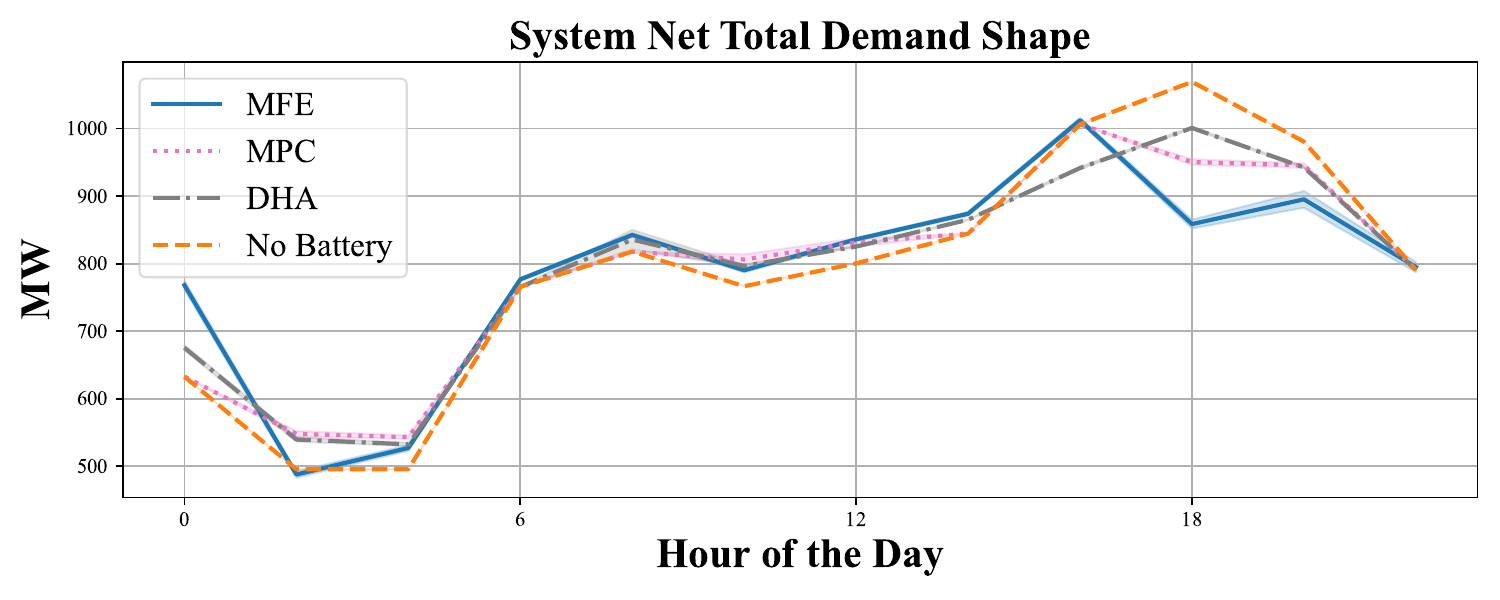}  
   \caption{Final demand profiles compared among the 3 algorithms and the scenario without battery storage. The vertical axis represents total sum energy demand across prosumers and consumers.}
    \label{fig:final-shape}
\end{figure}

One well-known challenge in power systems with high solar penetration is the so-callled `\textit{duck curve}' -- a net demand pattern characterized by steep ramping needs in the early evening when solar generation drops and consumption rises. This pattern strains grid flexibility and has become a major operational concern. 
Our numerical results demonstrate that the MFE-based learning framework effectively mitigates the duck curve issue. In Figure~\ref{fig:final-shape}, we illustrate this by comparing historical and reshaped net demand profiles with MFE and the 2 benchmark algorithms. The orange dashed line represents the historical net load from prosumers and pure consumers in Oahu, based on data from \cite{coffman2016estimating}. The remaining 3 curves show the reshaped aggregate demand of prosumers and consumers under the MFE policy and the two benchmark algorithms, averaged over the final 10 days of simulation. The MFE-based policy produces the most load shifting, with charging in low-price periods mostly around midnight and reducing net demand during peak evening hours. While the benchmark policies exhibit similar trends, their ability to reshape demand is more limited in comparison to the MFE case. This reshaping smooths the aggregate load profile and alleviates the duck curve effect. 

\section{Conclusion} \label{sec:conclusion}
In this paper, we engineer an algorithmic framework that prescribes how DERs can be incorporated into wholesale electricity markets, where decentralized aggregators employ automated RL algorithms to coordinate their participation. Specifically, our approach enables aggregators to intelligently manage energy storage charging and discharging on behalf of their prosumers under uncertainties from weather, demand, and market prices. The core idea is to treat the LMPs as a mean-field signal, which encapsulates aggregate supply-demand dynamics and network constraints. Within the proposed hybrid MFC-MFG framework, we introduce a two-phase RL-based learning algorithm that allows each aggregator to iteratively learn optimal policies in a decentralized manner.

We provide theoretical foundations by establishing sufficient conditions under which an MFE exists and is unique, and show that our three-step algorithmic approach converges to this equilibrium in an infinite-agent game. Numerical experiments demonstrate that incorporating storage with our RL-based framework significantly reduces LMP volatility, leading to a more stable market. Importantly, it also yields cost benefits for both prosumers and consumers, underscoring the broader value of market-driven DER integration. 

Future work will explore richer bidding strategies, including demand response, improved EV modeling, and simulations at the scale of large power systems such as CAISO or ERCOT. Another important direction is to study the strategic behavior of aggregators or VPPs that recognize their influence on LMPs and may manipulate prices to their advantage. Such behavior could disadvantage other market participants and undermine long-term market stability, for example, by driving prices too low, discouraging future capacity investments, and ultimately raising concerns about system reliability. A complementary theoretical direction is to analyze a finite-agent game and establish conditions under which its equilibria (if exist) converge to the mean-field equilibrium as the number of agents goes to infinity, thereby strengthening the link between finite-agent and mean-field models.

\appendices
\section{Proof Sketches}\label{app:Proof}

Due to space limitations, we provide only brief outlines of the main proofs and refer the reader to our arXiv preprint\footnote{Available at \url{https://arxiv.org/abs/2507.03240}} for complete details. 

\subsection{Uniqueness of Optimal Policy under Regularization}\label{app:uniqueness}
\begin{proposition}\label{prop:reg-unique}
    For any given LMP profile $\vlambda \in \RR^H$, there exists a unique optimal policy $\pi^*$ satisfying $\pi^* \in \arg\max_{\pi} V^\reg(s, \pi, \vlambda)$ for all $s \in \mathcal{S}$.
\end{proposition}
\begin{proof}
    Fix an arbitrary LMP profile $\vlambda \in \RR^H$. Since the state space $\mathcal{S}$ and action space $\mathcal{A}$ are finite, any value function $V(\cdot, \vlambda)$ can be identified as a vector $V \in \RR^{\abs{\mathcal{S}}}$, equipped with the supremum norm $\normof{V}_\infty = \max_{s \in \mathcal{S}} \abs{V(s)}$. For notational brevity, let $\pi^s_{\vlambda} := \pi(\cdot | s, \vlambda) \in \mathcal{P}(\mathcal{A})$ denote the action distribution at state $s$ under $\vlambda$, with probability $\pi^s_{\vlambda}(a) := \pi(a | s, \vlambda)$ for each $a \in \mathcal{A}$.

    For any stationary policy $\pi$ and vector $V \in \RR^{\abs{\mathcal{S}}}$, we define the regularized Bellman evaluation operator $\mathcal{T}^{\pi}_{\vlambda}: \RR^{\abs{\mathcal{S}}} \to \RR^{\abs{\mathcal{S}}}$ and the regularized Bellman optimality operator $\mathcal{T}^*_{\vlambda}: \RR^{\abs{\mathcal{S}}} \to \RR^{\abs{\mathcal{S}}}$ component-wise for each $s \in \mathcal{S}$ as 
    \begin{align*}
        & (\mathcal{T}^{\pi}_{\vlambda} V)(s) := \\
        &\hspace{5pt} \sum_{a \in \mathcal{A}} \pi^s_{\vlambda}(a) \left[ r(s, a, \vlambda) + \gamma \sum_{s' \in \mathcal{S}} P(s' | s, a) V(s') \right] - \Omega(\pi^s_{\vlambda}), \\
        &(\mathcal{T}^*_{\vlambda} V)(s) := \\
        &\hspace{5pt} \max_{\pi^s_{\vlambda} \in \mathcal{P}(\mathcal{A})} \Bigg\{ \sum_{a \in \mathcal{A}} \pi^s_{\vlambda}(a) \left[ r(s, a, \vlambda) + \gamma \sum_{s' \in \mathcal{S}} P(s' | s, a) V(s') \right] \\
        &\hspace{5pt}- \Omega(\pi^s_{\vlambda}) \Bigg\}.
    \end{align*}
    
    We first show that $\mathcal{T}^*_{\vlambda}$ is a $\gamma$-contraction with respect to $\normof{\cdot}_\infty$. For any $V_1, V_2 \in \RR^{\abs{\mathcal{S}}}$ and any $s \in \mathcal{S}$, using the property that $\abs{\max_{\pi^s_{\vlambda}} f(\pi^s_{\vlambda}) - \max_{\pi^s_{\vlambda}} g(\pi^s_{\vlambda})} \leq \max_{\pi^s_{\vlambda}} \abs{f(\pi^s_{\vlambda}) - g(\pi^s_{\vlambda})}$, the regularization term $\Omega(\pi^s_{\vlambda})$ and the immediate reward $r(s, a, \vlambda)$ cancel out, yielding
    \begin{align*}
        &\abs{(\mathcal{T}^*_{\vlambda} V_1)(s) - (\mathcal{T}^*_{\vlambda} V_2)(s)} \\
        &\leq \max_{\pi^s_{\vlambda}} \abs{ \gamma \sum_{a} \pi^s_{\vlambda}(a) \sum_{s'} P(s' | s, a) \big(V_1(s') - V_2(s')\big) } \\
        &\leq \gamma \max_{\pi^s_{\vlambda}} \sum_{a} \pi^s_{\vlambda}(a) \sum_{s'} P(s' | s, a) \normof{V_1 - V_2}_\infty \\
        &= \gamma \normof{V_1 - V_2}_\infty.
    \end{align*}
    Taking the maximum over $s$ establishes $\normof{\mathcal{T}^*_{\vlambda} V_1 - \mathcal{T}^*_{\vlambda} V_2}_\infty \leq \gamma \normof{V_1 - V_2}_\infty$. Because $\gamma \in [0, 1)$ and $(\RR^{\abs{\mathcal{S}}}, \normof{\cdot}_\infty)$ is a complete metric space, the Banach fixed-point theorem guarantees that $\mathcal{T}^*_{\vlambda}$ admits a unique fixed point $V^{\reg*}(\cdot, \vlambda) \in \RR^{\abs{\mathcal{S}}}$ satisfying $V^{\reg*}(\cdot, \vlambda) = \mathcal{T}^*_{\vlambda} V^{\reg*}(\cdot, \vlambda)$. By an identical argument, $\mathcal{T}^{\pi}_{\vlambda}$ is a $\gamma$-contraction with unique fixed point $V^\reg(\cdot, \pi, \vlambda)$, and standard monotonicity of the Bellman operators implies $V^{\reg*}(s, \vlambda) = \max_{\pi} V^\reg(s, \pi, \vlambda)$ for all $s$. We next establish the existence and uniqueness of the optimal policy $\pi^*$. Substitute the unique fixed point $V^{\reg*}(\cdot, \vlambda)$ into the definition of $\mathcal{T}^*_{\vlambda}$, and define the optimal $Q$-function as
    \begin{equation*}
        Q^{\reg*}(s, a, \vlambda) := r(s, a, \vlambda) + \gamma \sum_{s' \in \mathcal{S}} P(s' | s, a) V^{\reg*}(s', \vlambda).
    \end{equation*}
    By uniqueness of the fixed point of $\mathcal{T}^{\pi^*}_{\vlambda}$, a policy $\pi^*$ achieves $V^\reg(s, \pi^*, \vlambda) = V^{\reg*}(s, \vlambda)$ for all $s$ if and only if $\mathcal{T}^{\pi^*}_{\vlambda} V^{\reg*}(\cdot, \vlambda) = V^{\reg*}(\cdot, \vlambda) = \mathcal{T}^*_{\vlambda} V^{\reg*}(\cdot, \vlambda)$, which is equivalent to requiring that for every state $s$, the distribution $\pi^*$ solves the state-wise optimization problem
    \begin{equation}\label{eq:statewise_max}
        \max_{\pi^s_{\vlambda} \in \mathcal{P}(\mathcal{A})} \left\{ \sum_{a \in \mathcal{A}} \pi^s_{\vlambda}(a) Q^{\reg*}(s, a, \vlambda) - \Omega(\pi^s_{\vlambda}) \right\}.
    \end{equation}
    For each $s$, the probability simplex $\mathcal{P}(\mathcal{A})$ is a non-empty, compact, and convex subset of $\RR^{\abs{\mathcal{A}}}$. In~\eqref{eq:statewise_max}, the first term $\sum_{a} \pi^s_{\vlambda}(a) Q^{\reg*}(s, a, \vlambda)$ is linear in $\pi^s_{\vlambda}$, while $-\Omega(\pi^s_{\vlambda})$ is $\rho$-strongly concave by assumption. Consequently,~\eqref{eq:statewise_max} is $\rho$-strongly concave in $\pi^s_{\vlambda}$ over $\mathcal{P}(\mathcal{A})$ and admits a unique maximizer $\pi^*$ at every $s$. This uniquely defines the optimal policy $\pi^*$ across the entire state space.
\end{proof}

\subsection{Proof of Theorem \ref{thm:op}}
\label{app:pf-thm-op}
We first need the following property of Fenchel conjugates from Lemma 15 in \cite{shalev-fenchel}:
\begin{proposition}
Let $E=\RR^m$, $m\geq 1$, with inner product $\inner{\cdot}{\cdot}$, and let $g:E\to\RR\cup\{+\infty\}$ be proper, closed, and $\rho$-strongly convex with respect to a norm $\normof{\cdot}$. Define its Fenchel conjugate by $g^\star(y):=\sup_{x\in E}\{\inner{x}{y}-g(x)\}$. Then $g^\star$ is differentiable, $\nabla g^\star(y)$ is the unique maximizer of $\inner{x}{y}-g(x)$, and for any $y_1,y_2\in E$,
\begin{equation*}
\normof{\nabla g^\star(y_1)-\nabla g^\star(y_2)}\leq\frac{1}{\rho}\normof{y_1-y_2}_\star,
\end{equation*}
where $\normof{\cdot}_\star$ is the dual norm of $\normof{\cdot}$.
\label{prop:fc}
\end{proposition}
\begin{proof}
See Lemma 15 in \cite{shalev-fenchel}.
\end{proof}

We are now ready to present the proof of Theorem~\ref{thm:op}. Because the proof is symmetric for all aggregators, we fix an aggregator and drop its index except in $\Gamma_1^n$.
\begin{proof}[\textbf{Proof of Theorem \ref{thm:op}}]
For each LMP profile $\vlambda$, define the regularized Bellman optimality operator by
\begin{align*}
& (\mathcal T_{\vlambda}V)(s):=\max_{p\in\mathcal P(\mathcal{A})}\Big\{\sum_{a\in\mathcal{A}}p(a)\Big[r(s,a,\vlambda)\\
&\hspace{10pt}+\gamma\sum_{s'}P(s'\mid s,a)V(s')\Big]-\Omega(p)\Big\}.
\end{align*}
This operator is a $\gamma$-contraction in the supremum norm, so it has a unique fixed point $V^\star_{\vlambda}$. Define the action-score function and its statewise vector by
\begin{equation}
Q_{\vlambda}(s,a):=r(s,a,\vlambda)+\gamma\sum_{s'}P(s'\mid s,a)V^\star_{\vlambda}(s'),
\label{eq:Q}
\end{equation}
\begin{equation}
q_{\vlambda}^s:=\bigl(Q_{\vlambda}(s,a)\bigr)_{a\in\mathcal{A}}.
\label{eq:q}
\end{equation}
The Bellman fixed-point equation can therefore be written as
\begin{equation}
V^\star_{\vlambda}(s)=\max_{p\in\mathcal P(\mathcal{A})}\left\{\inner{p}{q_{\vlambda}^s}-\Omega(p)\right\}=\Omega^\star(q_{\vlambda}^s).
\label{eq:Qip}
\end{equation}

We first bound the change in the optimal value when the LMP profile changes. For any fixed $V$, compare the maximization defining $\mathcal T_{\vlambda_1}V$ and $\mathcal T_{\vlambda_2}V$ using the same distribution $p$. The regularization terms then cancel, and~\eqref{eqn:reward-lipschitz} gives $\normof{\mathcal T_{\vlambda_1}V-\mathcal T_{\vlambda_2}V}_\infty\leq L_r\normof{\vlambda_1-\vlambda_2}_1$. Since each Bellman operator is a $\gamma$-contraction,
\begin{align*}
&\normof{V^\star_{\vlambda_1}-V^\star_{\vlambda_2}}_\infty\\
&\leq\normof{\mathcal T_{\vlambda_1}V^\star_{\vlambda_1}-\mathcal T_{\vlambda_1}V^\star_{\vlambda_2}}_\infty
+\normof{\mathcal T_{\vlambda_1}V^\star_{\vlambda_2}-\mathcal T_{\vlambda_2}V^\star_{\vlambda_2}}_\infty\\
&\leq\gamma\normof{V^\star_{\vlambda_1}-V^\star_{\vlambda_2}}_\infty
+L_r\normof{\vlambda_1-\vlambda_2}_1.
\end{align*}
Rearranging yields
\begin{equation}
\normof{V^\star_{\vlambda_1}-V^\star_{\vlambda_2}}_\infty
\leq\frac{L_r}{1-\gamma}\normof{\vlambda_1-\vlambda_2}_1.
\label{eq:L0}
\end{equation}

Next, the definition of $q_{\vlambda}^s$, the reward bound, and~\eqref{eq:L0} imply, for every state $s$,
\begin{align}
&\normof{q_{\vlambda_1}^s-q_{\vlambda_2}^s}_\infty \nonumber\\
&=\max_{a\in\mathcal{A}}\Big\vert r(s,a,\vlambda_1)-r(s,a,\vlambda_2)\nonumber\\
& \hspace{10pt}+\gamma\sum_{s'}P(s'\mid s,a)(V^\star_{\vlambda_1}(s')-V^\star_{\vlambda_2}(s') )\Big\vert \nonumber\\
&\leq L_r\normof{\vlambda_1-\vlambda_2}_1+\gamma\normof{V^\star_{\vlambda_1}-V^\star_{\vlambda_2}}_\infty \nonumber\\
&\leq\frac{L_r}{1-\gamma}\normof{\vlambda_1-\vlambda_2}_1. \label{eq:q_Lip}
\end{align}
Finally, apply Proposition~\ref{prop:fc} to $\Omega$. Since $\Omega$ includes the constraint $p\in\mathcal P(\mathcal{A})$, its conjugate gradient is precisely the optimal action distribution at state $s$:
\begin{equation*}
\Gamma_1^n(s,\vlambda)=\nabla\Omega^\star(q_{\vlambda}^s).
\end{equation*}
The dual of the $\ell_1$ norm is the $\ell_\infty$ norm. Thus, uniformly over $s\in\mathcal S$,
\begin{equation}
\normof{\Gamma_1^n(s,\vlambda_1)-\Gamma_1^n(s,\vlambda_2)}_1
\leq\frac{1}{\rho}\normof{q_{\vlambda_1}^s-q_{\vlambda_2}^s}_\infty
\leq L_1\normof{\vlambda_1-\vlambda_2}_1.
\label{eq:Gamma_Lip}
\end{equation}
where $L_1 = \frac{L_r}{\rho(1-\gamma)}$. This concludes the proof.
\end{proof}

\subsection{Proof of Theorem \ref{thm:gamma2}}
\label{app:pf-thm-co}
\begin{proof}
We expand the no-regeneration transition operator as
\begin{equation*}
\tilde{\Gamma}_2^n(\LL^n,\pi^n)(s',a') = \sum_{s,a}\LL^n(s,a) P^n(s'|s,a) \pi^n(a'|s').
\end{equation*}
The LMP profile is fixed throughout and is omitted in this proof's notation for simplicity. Let $U^n(s')$ be the next-state distribution conditional on the uniform SOC reset in~\eqref{eqn:random-regen}. Let $\overline{\zeta}=1-\zeta$. The consistency operator is
\begin{equation*}
\Gamma_2^n(\LL^n,\pi^n)(s',a') = \zeta U^n(s')\pi^n(a'|s')+\overline{\zeta}\tilde{\Gamma}_2^n(\LL^n,\pi^n)(s',a').
\end{equation*}
First, fix $\pi^n$ and compare $\LL^n$ and $\LL^{n'}$ at time $t$. The reset terms are identical and cancel. By the triangle inequality and because $P^n$ and $\pi^n$ are probability kernels, we get
\begin{align*}
    & \normof{\Gamma_2^n(\LL^n,\pi^n)-\Gamma_2^n(\LL^{n'},\pi^n)}_1 \\
    &=\overline{\zeta}\sum_{s',a'}\abs{\sum_{s,a}\bigl(\LL^n(s,a)-\LL^{n'}(s,a)\bigr)
      P^n(s'|s,a)\pi^n(a'|s')}\\
    &\leq\overline{\zeta}\sum_{s,a}\abs{\LL^n(s,a)-\LL^{n'}(s,a)}\sum_{s'}P^n(s'|s,a)\sum_{a'}\pi^n(a'|s')\\
    &=\overline{\zeta}\normof{\LL^n-\LL^{n'}}_1.
\end{align*}
Thus the first bound holds with $L_2=\overline{\zeta}=1-\zeta$. Next, fix $\LL^n$ and compare two policies $\pi^n$ and $\pi^{n'}$. Since $\LL^n$ is a probability distribution,
\begin{align*}
    &\normof{\Gamma_2^n(\LL^n,\pi^n)-\Gamma_2^n(\LL^n,\pi^{n'})}_1\\
    &=\sum_{s',a'}\Big\vert \sum_{s,a}\LL^n(s,a)\bigl[\zeta U^n(s')+\overline{\zeta}P^n(s'|s,a)\bigr]\\
    & \hspace{10pt} \cdot \bigl(\pi^n(a'|s')-\pi^{n'}(a'|s')\bigr)\Big\vert\\
      &\leq\sum_{s,a}\LL^n(s,a)\sum_{s'}\bigl[\zeta U^n(s')+\overline{\zeta}P^n(s'|s,a)\bigr]\\
    & \hspace{10pt} \cdot \sum_{a'}\abs{\pi^n(a'|s')-\pi^{n'}(a'|s')}\\
    &\leq\sup_{s'\in\mathcal S}\normof{\pi^n(\cdot|s')-\pi^{n'}(\cdot|s')}_1.
\end{align*}
Thus the second bound holds with $L_3=1$.
\end{proof}

\subsection{Proof of Theorem \ref{thm:mfe}}
\label{app:pf-mfe}
\begin{proof}
Let $\mathbf L=(\vLL_1,\ldots,\vLL_H)$ and $\widetilde{\mathbf L}=(\widetilde{\vLL}_1,\ldots,\widetilde{\vLL}_H)$ be two daily MF profiles, where $\vLL_h=(\LL_h^1,\ldots,\LL_h^N)$. Define $\normof{\mathbf L-\widetilde{\mathbf L}}_1:=\sum_{h=1}^H\sum_{n=1}^N\normof{\LL_h^n-\widetilde{\LL}_h^n}_1$.

Fix an hour $h$ and hold pure-consumer bids and other external market inputs fixed. For bus $n$, let $g^n(s,a):=\Phi(e,a,\eta_n)+q$ and $g_{\max}^n:=\max_{s,a}\abs{g^n(s,a)}$, where $q$ is the exogenous net-load component of $s$. By the capacity-weighted definition of $\LL_h^n$, the bid is
\begin{equation*}
D_h^n(\LL_h^n)=D_h^{n,\mathrm c}+\overline E^n\sum_{s,a}\LL_h^n(s,a)g^n(s,a).
\end{equation*}
The pure-consumer term cancels when the two MF profiles are compared. Therefore,
\begin{align*}
\abs{D_h^n(\LL_h^n)-D_h^n(\widetilde{\LL}_h^n)}
&=\overline E^n\abs{\sum_{s,a}\bigl(\LL_h^n(s,a)-\widetilde{\LL}_h^n(s,a)\bigr)g^n(s,a)}\\
&\leq\overline E^n g_{\max}^n\normof{\LL_h^n-\widetilde{\LL}_h^n}_1.
\end{align*}
Summing over demand buses yields
\begin{align*}
\normof{\mD_h(\vLL_h)-\mD_h(\widetilde{\vLL}_h)}_1
&\leq\sum_{n=1}^N\overline E^n g_{\max}^n\normof{\LL_h^n-\widetilde{\LL}_h^n}_1\\
&\leq\max_n\{\overline E^n g_{\max}^n\}\normof{\vLL_h-\widetilde{\vLL}_h}_1.
\end{align*}
Applying Proposition~\ref{prop:feng2025}, with $L_\lambda$ defined for MWh demand bids, gives for every price bus $n$
\begin{equation*}
\abs{\lambda_h^n(\vLL_h)-\lambda_h^n(\widetilde{\vLL}_h)}\leq L_{\mathrm{MF}}\normof{\vLL_h-\widetilde{\vLL}_h}_1,
\end{equation*}
where $L_{\mathrm{MF}}:=L_\lambda\max_n\{\overline E^n g_{\max}^n\}$. Let $\vlambda^n(\mathbf L)$ be the $H$-entry price profile induced at bus $n$ by $\mathbf L$. Summing the previous bound over hours gives
\begin{align*}
\normof{\vlambda^n(\mathbf L)-\vlambda^n(\widetilde{\mathbf L})}_1
&=\sum_{h=1}^H\abs{\lambda_h^n(\vLL_h)-\lambda_h^n(\widetilde{\vLL}_h)}\\
&\leq L_{\mathrm{MF}}\normof{\mathbf L-\widetilde{\mathbf L}}_1.
\end{align*}

Let $\pi_{\mathbf L}^n$ be the exact optimal policy for $\vlambda^n(\mathbf L)$, defined at every state by $\Gamma_1^n$. Define the exact daily MF map by $\Gamma(\mathbf L)_{h+1}^n:=\Gamma_{2,h}^n(\LL_h^n,\pi_{\mathbf L}^n)$, with $h+1$ interpreted cyclically after hour $H$ and $\Gamma_{2,h}^n$ denoting the hour-$h$ consistency update. The Lipschitz bounds for the consistency and optimality operators imply
\begin{align*}
&\normof{\Gamma(\mathbf L)_{h+1}^n-\Gamma(\widetilde{\mathbf L})_{h+1}^n}_1\\
&\leq L_2\normof{\LL_h^n-\widetilde{\LL}_h^n}_1+L_3\sup_{s\in\mathcal S}\normof{\pi_{\mathbf L}^n(\cdot\mid s)-\pi_{\widetilde{\mathbf L}}^n(\cdot\mid s)}_1\\
&\leq L_2\normof{\LL_h^n-\widetilde{\LL}_h^n}_1+L_1L_3\normof{\vlambda^n(\mathbf L)-\vlambda^n(\widetilde{\mathbf L})}_1.
\end{align*}
Summing over hours and buses, we obtain
\begin{align*}
&\normof{\Gamma(\mathbf L)-\Gamma(\widetilde{\mathbf L})}_1\\
&\leq L_2\normof{\mathbf L-\widetilde{\mathbf L}}_1+HL_1L_3\sum_{n=1}^N\normof{\vlambda^n(\mathbf L)-\vlambda^n(\widetilde{\mathbf L})}_1\\
&\leq\left(L_2+HNL_1L_3L_{\mathrm{MF}}\right)\normof{\mathbf L-\widetilde{\mathbf L}}_1.
\end{align*}
Thus the exact daily map is a contraction when $L_2+HNL_1L_3L_{\mathrm{MF}}<1$. Banach's fixed-point theorem gives a unique fixed point.
\end{proof}

\bibliographystyle{IEEEtran}
\bibliography{references}
\end{document}